\documentclass[aps,prb,superscriptaddress,showpacs,showkeys,floatfix,reprint]{revtex4-1}
\usepackage{graphicx}
\usepackage{dcolumn}
\usepackage[pdftex]{color}
\usepackage{amsmath}
\usepackage{caption}
\usepackage{subcaption}
\usepackage{bm}
\usepackage{gensymb}
\usepackage{verbatim}
\definecolor{darkgreen}{rgb}{0.133,0.545,0.133}
\definecolor{orange}{rgb}{1.0,0.76,0.02}

\begin{document}

\title{Phase transitions in titanium with an analytic bond-order potential}

\author{Alberto Ferrari}
\email{alberto.ferrari@rub.de}
\affiliation{Interdisciplinary Centre for Advanced Materials Simulation, Ruhr-Universit{\"a}t Bochum, 44801 Bochum, Germany}
\author{Malte Schr\"oder}
\affiliation{Interdisciplinary Centre for Advanced Materials Simulation, Ruhr-Universit{\"a}t Bochum, 44801 Bochum, Germany}
\author{Yury Lysogorskiy}
\affiliation{Interdisciplinary Centre for Advanced Materials Simulation, Ruhr-Universit{\"a}t Bochum, 44801 Bochum, Germany}
\author{Jutta Rogal}
\affiliation{Interdisciplinary Centre for Advanced Materials Simulation, Ruhr-Universit{\"a}t Bochum, 44801 Bochum, Germany}
\author{Matous Mrovec}
\affiliation{Interdisciplinary Centre for Advanced Materials Simulation, Ruhr-Universit{\"a}t Bochum, 44801 Bochum, Germany}
\author{Ralf Drautz}
\affiliation{Interdisciplinary Centre for Advanced Materials Simulation, Ruhr-Universit{\"a}t Bochum, 44801 Bochum, Germany}

\date{\today}

\begin{abstract}
Titanium is the base material for a number of technologically
important alloys for energy conversion and structural applications.
Atomic-scale studies of Ti-based metals employing first-principles
methods, such as density functional theory, are limited to
ensembles of a few hundred atoms.  To perform large-scale and/or
finite temperature simulations, computationally more efficient
interatomic potentials are required. In this work, we
coarse grain the tight-binding (TB) approximation to the electronic
structure and develop an analytic bond-order potential (BOP) for Ti by
fitting to the energies and forces of elementary deformations of
simple structures.  The BOP predicts the structural properties
of the stable and defective phases of Ti with a quality comparable to
previous TB parametrizations at a much lower computational cost. The
predictive power of the model is demonstrated for simulations of martensitic transformations.
\end{abstract}

%\pacs{88.88.Aa, 88.88.Aa, 88.88.Aa}

%\keywords{Ti-Ta}

\maketitle

\section{Introduction}

%\textcolor{green}{Parts if we decide to keep the pressure-induced transformations.}
Titanium alloys are very attractive materials for structural and
functional applications, superior to steels concerning the
stiffness-to-weight and strength-to-weight ratios, corrosion
resistance, and biocompatibility \cite{Ashby1998}. Ti-based materials are
also characterized by unique elastic and mechanical properties, such
as the shape memory alloys Ti-Ni \cite{Buehler1963}, Ti-Pd, Ti-Pt,
Ti-Au \cite{Doonkersloot1970}, Ti-Nb \cite{Kim2006}, Ti-Ta
\cite{Bagaryatskii1958,Ferrari2019} and Ti-Mo
\cite{Endoh2017}, or the gum metals Ti-Nb-Ta-Zr-O and Ti-Ta-Nb-V-Zr-O
\cite{Saito2003}.

The remarkable properties of Ti alloys descend from the rich phase
diagram of this element: Ti has five thermodynamically stable solid
phases, the $\alpha$ phase (hcp, spacegroup $P6_{3}/mmc$,
\textit{Strukturbericht} designation A3), the $\beta$ phase (bcc, $Im\bar{3}m$, A2), the
$\omega$ phase (hexagonal, $P6/mmm$, C32), and the $\gamma$
\cite{Vohra2001} and $\delta$ \cite{Akahama2001} phases (both
orthorhombic, $Cmcm$, A20).  At room temperature and ambient pressure, Ti is hcp and
transforms martensitically to bcc at high temperatures and to $\omega$,
$\gamma$, and $\delta$ at high pressures. At zero temperature and
pressure, not accessible to experiments,
there is a general consensus that the ground state is the $\omega$
phase \cite{Trinkle2003}, which is more stable than hcp, $\gamma$ and
$\delta$ by less than 10~meV/at.~and bcc by more than 100~meV/at.

Atomistic investigations of fundamental structural and thermodynamic
properties of Ti-based materials are commonly performed using density
functional theory (DFT). However, DFT calculations are limited to small
length- ($<$ 5 nm$^{3}$) and time- ($<$ 10 ps) scales and therefore direct
studies of extended defects or phase
transitions are usually not possible. To carry out such simulations, several
empirical potentials have been fitted to experimental or first-principles
data. These potentials are generally classical potentials based on the
embedded atom method (EAM) \cite{Daw1984} or modified embedded atom method
(MEAM) \cite{Baskes1992}.  Such empirical models are unable to fully
capture subtle features of the mixed metallic-covalent bonding in Ti and this often leads
to quantitative or even qualitative discrepancies in the predicted properties
of some phases.  For instance, it has been reported that an accurate description
of the $\omega$ phase \cite{Ackland1992, Zope2003, Zhou2004, Ko2015,
  Gibson2016, Dickel2018} or of the temperature-dependent behaviour of the bcc phase
\cite{Mendelev2016,Kartamyshev2019} needs to be sacrificed to  achieve an
overall good accuracy of the potential.  A few potentials have succeeded to reproduce
at least the $\alpha$, $\beta$, and $\omega$ phases quantitatively 
\cite{Hennig2008, Ehemann2017, Takahashi2017} by increasing the model complexity and by
employing non-smooth interpolators, which however might lead
to overfitting. The transferability of these more complex
potentials to properties or environments not included in the training has been
questioned \cite{Gibson2016, Rawat2017}.

An alternative to DFT and classical potentials are tight-binding (TB)
models \cite{Slater1954, Ashcroft1976, Mehl1996, Mehl2002, Rudin2004, Trinkle2003, Trinkle2006, Margine2011, Cawkwell2015}.
Nonorthogonal TB models with $spd$-basis
have been proven successful in the description of the most relevant
properties of the hcp, bcc, $\omega$, and $\gamma$ phases in Ti
\cite{Mehl2002,Trinkle2006}. These TB models contain more than 100 parameters and their parameterization is often
elaborate.  In addition, similiarly to DFT, the diagonalization of the
Hamiltonian and overlap matrices results in an unfavorable cubic $O(N^3)$
scaling of the computational cost with the number of
atoms. Hence, there is a demand for models that capture the
essential characteristics of the electronic structure of Ti but are
computationally efficient and simple to construct. One of such schemes
are bond-order potentials (BOPs) \cite{Pettifor1989,Horsfield1996,Pettifor2002,Drautz2006}, linear scaling interatomic
potentials derived by coarse graining the TB method. 
Unfortunately, the only BOP for Ti in the literature \cite{Girshick1998, Girshick1998a} 
fails to  accurately reproduce crucial properties of this element, including the cohesive energies of the $\omega$, fcc, and bcc phases, 
because its parametrization focused mainly on the hcp structure.  

In this work, we develop a new, simple $d$-valent analytic BOP for Ti and solve some of the critical flaws of the previous BOP for Ti. Our
model contains only 25 adjustable parameters fitted to DFT energies of
elementary structures at 0 K conditions. Despite its simplicity, our BOP
accurately describes  the main features of the bonding in Ti: it shows
a good transferability to atomic environments not included in the fit
set, qualitative agreement with first principles calculations on
high-pressure and defective structures, and
quantitative agreement with experiments regarding finite temperature
properties.

This article is organized as follows: Sec.~\ref{methods} introduces
the theoretical background, the level of approximation and the
structure of our model for Ti.  Sec.~\ref{fit} describes the fitting
strategy with the database of fitted quantities  and basic validation tests.
Sec.~\ref{simple_phases} contains the data on simple structures at 0
K.  Sec.~\ref{defects} studies the defect properties.
Sec.~\ref{temperature} presents the tests of the potential on the
phase transitions induced by temperature and pressure before we
conclude our work in Sec.~\ref{sec:conclusions}.

\section{Methodology}
\label{methods}

Ti has four valence electrons; formally, two of these electrons have
an $s$-character and the remaining two a $d$-character. To maximize
the completeness of the TB representation, TB models for Ti usually
employ a full nonorthogonal basis set that contains $s$, $p$, and $d$
angular components (a nonorthogonal \textit{spd} model). This means
that ten bond and ten overlap integrals ($ss\sigma$, $sp\sigma$,
$pp\sigma$, $pp\pi$, $sd\sigma$, $pd\sigma$, $pd\pi$, $dd\sigma$,
$dd\pi$, $dd\delta$ in the Slater-Koster notation \cite{Slater1954})
have to be parametrized as a function of the interatomic distance,
making the fitting procedure of these models rather cumbersome.

Nonorthogonal \textit{$spd$} models can be simplified in three
conceptual steps. In a first step, at the cost of introducing an
environmental dependence of the bond integrals \cite{Nguyen-Manh2000}, the number of fitting parmeters can be halved
by considering an \textit{orthogonal} TB model, which can be derived
from a nonorthogonal one by applying, for instance, L\"owdin symmetric
orthogonalization \cite{Loewdin1950,Urban2011}:
\begin{equation}
\tilde{H}=S^{-\frac{1}{2}}HS^{\frac{1}{2}} \quad ,
\end{equation}
where $\tilde{H}$ and $H$ are the Hamiltonian matrices of orthogonal and
nonorthogonal models, respectively, and $S$ is the overlap matrix in the
nonorthogonal model. In a second step, the explicit treatment of interactions
between orbitals with $s$- and $p$-characters can be neglected, since the
unsaturated directional bonds governing the structural stability of transition
metals originate predominantly from the $d$-electron interactions \cite{Pettifor1970}.
As described in detail below, an
orthogonal $d$-only model is sufficient to correctly capture the small
energy differences between the most stable phases of Ti. Finally, in a third
step, the TB model can be coarse-grained to a BOP.

\subsection{Significance of an orthogonal $d$-only model}
\label{canonical}

\begin{figure}
\begin{centering} 
\includegraphics[scale=0.16]{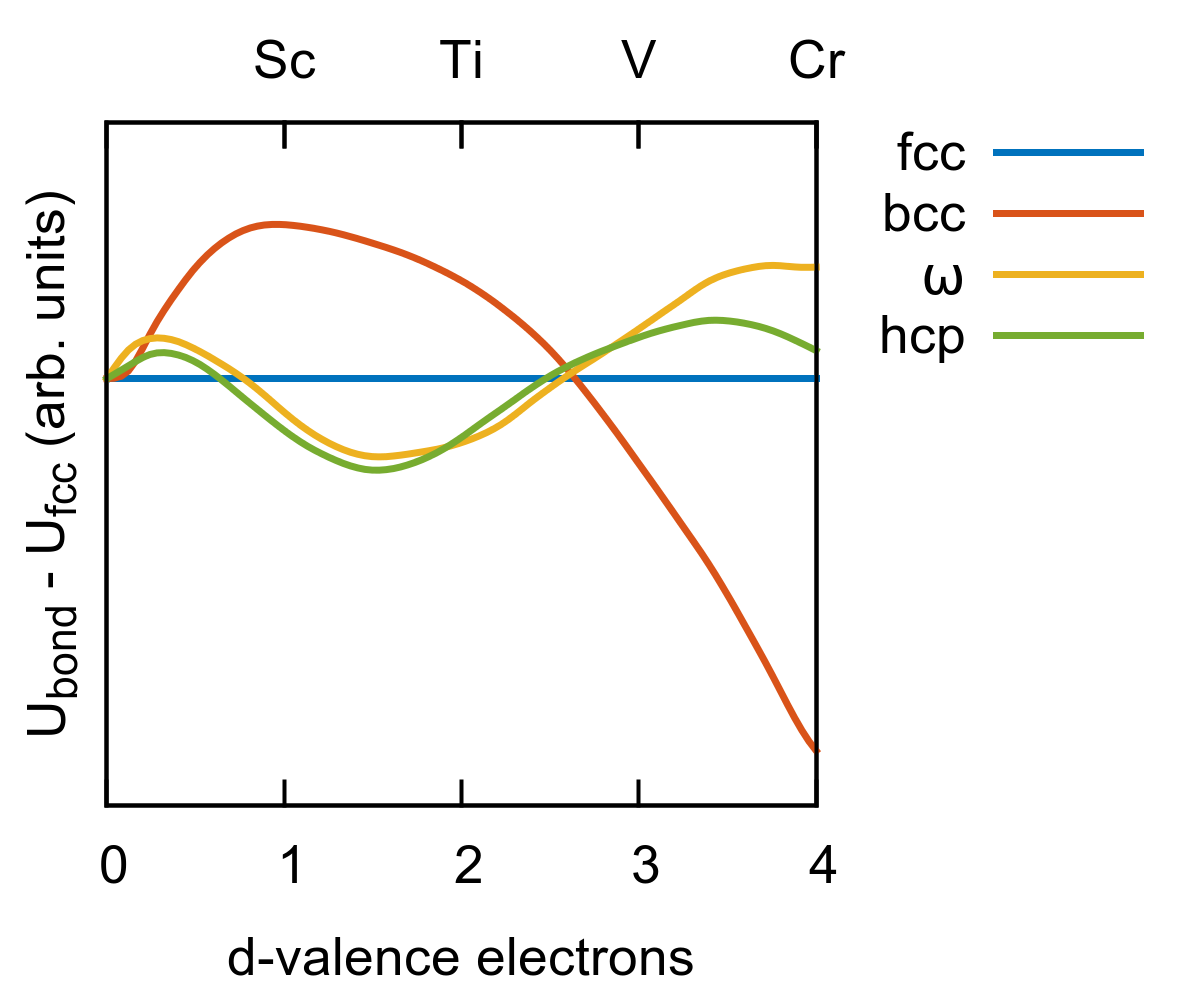}
\par\end{centering}
\caption{Bonding energy difference with respect to the fcc structure
  as a function of the $d$-band filling for the structures fcc, bcc,
  $\omega$, and hcp. The volume of the considered phases was
  adjusted so that their repulsion contributions were the same.}
\label{struc_diff_th}
\end{figure}

To prove that the relative stability of the thermodynamically stable phases of
Ti can be captured by an orthogonal $d$-only model, we employed the
structural energy difference theorem \cite{Pettifor1986, Seiser2011}.  At low
pressure the orthorhombic phases $\gamma$ and $\delta$ are degenerate with the
hcp structure, hence we focused only on the hcp, bcc, and $\omega$ structures.
The binding energy $U$ of a given atomic configuration is the sum of a
bonding term $U_\text{bond}$, in this case due to the $d$-electrons only, and a
repulsive term $U_\text{rep}$:
\begin{equation}
U=U_\text{bond}+U_\text{rep} \quad .
\end{equation}
The structural energy difference theorem states that, to first
order, the binding energy difference between two structures at
equilibrium distance is
\begin{equation}
\Delta U^{(1)}=\left[\Delta U_\text{bond}\right]_{\Delta U_\text{rep}=0} \quad ,
\end{equation}
that is, the relative stability can be evaluated by comparing the
bonding terms of the two structures, provided that the repulsive
contributions of the two structures are the same.

Here, for a qualitative description of $U_\text{bond}$, we chose
an orthogonal $d$-only model with the $d$ bond integrals having
the canonical form \cite{Pettifor1977, Andersen1978}
\begin{equation}
\left.\begin{array}{c}
dd\sigma\\
dd\pi\\
dd\delta
\end{array}\right\} =\left.\begin{array}{c}
-6\\
4\\
-1
\end{array}\right\} \cdot\frac{C}{r^{5}} \quad ,
\end{equation}
where $r$ is the interatomic distance and $C$ is a constant that
has the unit of energy.
Only the contributions from the first nearest neighbor shells for fcc, $\omega$ and hcp, and the first two nearest neighbor shells for bcc were considered.

$U_\text{rep}$ is given as a sum of a pairwise repulsion between the atoms
\begin{equation}
U_\text{rep}=\sum_{i\neq j}\phi\left(r_{ij}\right) \quad .
\label{repulsion}
\end{equation}
For an approximate evaluation of the relative phase stability,
$\phi\left(r\right)$ may be assumed to be dominated by the overlap
contribution and thus proportional to the square of the bond integrals
\cite{Pettifor1995, Seiser2011}
\begin{equation}
\phi\left(r\right) \propto \frac{1}{r^{10}} \quad .
\end{equation}
Using this simplified model, we varied the volume per atom of bcc,
hcp, $\omega$ and fcc phases to ensure their repulsive contributions
were equal and compared $U_\text{bond}$ for all four phases.
Fig.~\ref{struc_diff_th} shows $U_\text{bond}-U_\text{bond}^\text{(fcc)}$ as a
function of the $d$-band filling (the number of $d$ electrons).  For a
band filling of roughly 2.0-2.3 $d$ electrons, corresponding to Ti,
Zr, and Hf, the canonical bond integrals predict the correct ordering
of the most important phases in Ti, with the $\omega$ phase slightly
more stable than hcp, and bcc considerably higher in energy.  This
means that a simple orthogonal $d$-only model does provide the correct
phase stability in Ti.

\subsection{Bond-order potentials}

Given that the $d$-valence electrons are sufficient to take into account the
phase ordering in Ti, we aimed for a $d$-only model to develop our interatomic
potential. Instead of a TB model, we parametrized a
more computationally efficient BOP.  Besides the already mentioned potential for titanium \cite{Girshick1998, Girshick1998a}, $d$-valent BOPs have been proven very
successfull for many other transition metals, including molybdenum \cite{Mrovec2004,
  Cak2014, Lin2014}, iridium \cite{Cawkwell2005,Cawkwell2006}, tungsten
\cite{Mrovec2007, Cak2014, Lin2014}, iron \cite{Mrovec2011, Lin2016},
niobium, tantalum \cite{Cak2014, Lin2014}, and manganese \cite{Drain2014}.

BOPs are linear scaling quantum-mechanical potentials that retain
information on the TB electronic structure via the moments of the
local density of states (DOS). The $N$-th moment of the DOS $n\left(E\right)$
of the orbital $\alpha$ on atom $i$ is
\begin{equation}
\mu_{i\alpha}^{(N)}=\int E^N n_{i\alpha}\left(E\right)dE \quad .
\end{equation}
The moments of the DOS are related to the crystal structure via the moments theorem\cite{Cyrot-Lackmann1967,Jenke2018}, which links the $N$-th moment to a self-returning hopping path of length $N$ starting and ending on the orbital $\left|i\alpha\right\rangle$, assuming an orthonormal basis,
\[
\mu_{i\alpha}^{(N)}=\langle i\alpha|\hat{H^N}|i\alpha \rangle = \sum_{j\beta,k\gamma,...}\langle i\alpha|\hat{H}|j\beta\rangle 
\]
\begin{equation}
\times\langle j\beta|\hat{H}|k\gamma\rangle \langle k\gamma|\hat{H}|...\rangle \langle ...|\hat{H}|i\alpha\rangle \quad .
\label{self_ret_path}
\end{equation}

If only the first
$N_\text{max}$ moments are considered, the local DOS, total energy, and
forces can be calculated analytically from the self-returning paths of lengths $\leq N_\text{max}$ at a
computational cost that scales linearly with the number of atoms in
the simulation cell~\cite{Pettifor1989, Horsfield1996, Drautz2006, Seiser2013}. BOPs thus offer a
great computational advantage over orthogonal TB models with only a
minor sacrifice in accuracy related to the truncation of the moments
expansion.

In this work we employed analytic BOPs \cite{Drautz2006,Drautz2011,Seiser2013}.
All BOP calculations were performed
using the BOPfox code \cite{Hammerschmidt2018} with a value of
$N_\text{max}=9$ and a terminator of 200 constant recursion coefficients.

\subsection{The BOP model for Ti}

The binding energy in our BOP model is expressed as
%5
\begin{equation}
U=U_\text{bond}+U_\text{emb}+U_\text{rep} \quad .
\end{equation}
The bonding energy $U_\text{bond}$ depends on the $dd\sigma$, $dd\pi$, and
$dd\delta$ two-center bond integrals. We modeled the distance dependence of
the bond integrals with the sum of two exponential functions:
\begin{equation}
\beta\left(r\right)=a_{1}e^{-b_{1}r^{c_{1}}}+a_{2}e^{-b_{2}r^{c_{2}}} \quad ,
\end{equation}
where $a_{i}$, $b_{i}$, and $c_{i}$
are adjustable parameters. The bond integrals were multiplied by a cutoff function,
\begin{equation}
f_\text{cut}\left(r\right)=\frac{1}{2}\left[\cos\left(\pi\frac{r-r_\text{cut}+d_\text{cut}}{d_\text{cut}}\right)+1\right] \quad ,
\end{equation}
in the range $r_\text{cut}-d_\text{cut}\leq r\leq r_\text{cut}$ to ensure their smooth decay
to zero. Values of 4.45 \AA~ and 1.35 \AA~were chosen for $r_\text{cut}$ and
$d_\text{cut}$, respectively.

Following Madsen \textit{et al.} \cite{Madsen2011} and Drain\textit{ et al.}
\cite{Drain2014}, an embedding function was introduced to mimic the
contribution of the missing $s$ electrons and $sd$ hybridization to the
cohesive energy.  The embedding term was parameterized using a Finnis-Sinclair
\cite{Finnis1984} second-moment expression:
\begin{equation}
U_\text{emb}=-\sum_{i}\sqrt{\sum_{j\neq i}\rho\left(r_{ij}\right)} \quad ,
\end{equation}
where $\rho\left(r\right)$ is represented by a smooth third-order spline
function with only two nodes.  Albeit empirical, the embedding term mimics the
bonding contribution of $s$ electrons acting as a homogeneous gas of
nearly-free electrons with density $\rho(r)$, in direct analogy to the EAM
potentials.

Finally, the repulsive term $U_\text{rep}$, including the overlap, electrostatic,
exchange-correlation, and double counting contributions, was parameterized
using a pairwise expression (Equation~\eqref{repulsion}), where
$\phi\left(r\right)$ is an exponential function with three fitting parameters
\begin{equation}
\phi\left(r\right)=Ae^{-Br^{C}} \quad .
\end{equation}
Motivated by the qualitative results obtained with the canonical model
(Figure \ref{struc_diff_th}), we fixed the number of $d$ electrons in our BOP
to 2.1. Changes in the number of electrons in the range 2.0-2.7 followed
by refitting did not improve the quality of the interatomic potential.

\section{Fitting strategy}
\label{fit}

\subsection{Fitting database}

Our fitting database \cite{Lysogorskiy2019} consisted of high-quality DFT energies and forces for
different atomic configurations calculated using the Vienna Ab initio
Simulation Package (VASP 5.4) \cite{Kresse1993, Kresse1996, Kresse1996a}
following the \textit{pyiron} workflow \cite{Janssen2018}. For all DFT
calculations we used a projector-augmented wave (PAW) pseudopotential
\cite{Bloechl1994,Kresse1999} with 12 valence electrons and the
exchange-correlation potential with the generalized-gradient expression by
Perdew, Burke, and Ernzerhof (PBE) \cite{Perdew1996}.  Standard values of 500
eV and 0.1 $2\pi/$\AA~were employed for the energy cutoff and $k$-point
linear density, respectively, to minimize the numerical errors. The $k$-points
were distributed in the Brillouin zone with the Monkhorst-Pack
\cite{Baldereschi1973,Monkhorst1976} special $k$-point technique.  The
electronic occupations were smeared using the Methfessel-Paxton function of
order one \cite{Methfessel1989} with a width of 0.2 eV.

The ground state energies of the $\omega$, hcp, double-hcp (dhcp), fcc, bcc, and A15 structures
were carefully determined by optimizing the volume and the $c/a$ ratio in the
hexagonal structures to ensure a precision of the obtained energies within less than
1 meV/at. Typically, the optimizations were done in two or three stages to
minimize the Pulay stresses \cite{DaCosta1986}.

Furthermore, the equilibrium cell of each phase was deformed
isotropically within a $\pm10\%$ volume range to obtain energy-volume
curves. For the $\omega$, hcp, and bcc phases we evaluated also the
elastic constants, by considering a series of symmetrically inequivalent
deformations ($\leq$0.5\%) \cite{Golesorkhtabar2013,Lysogorskiy2019}, and the phonon spectra, using the small
displacements method as implemented in Phonopy \cite{Togo2015}.

The BOP binding energy per atom $U$ is related to the DFT
energy per atom of the pseudopotential calculation $E$ by
\begin{equation}
U=E-E_\text{at} \quad ,
\end{equation}
where $E_\text{at}$ is the total energy of isolated Ti atoms. Since our BOP model
is constructed with respect to a non-magnetic Ti atom with electronic
configuration $\left[Ar\right]3s^{2}3d^{2}$, we took
$E_\text{at}$ as the energy of a spin-unpolarized Ti atom with the same electronic
configuration. For the pseudopotential used in this work, we obtained
$E_\text{at}=-1.16$ eV.

\subsection{Fitting procedure}

\begin{figure}
\begin{centering} 
\includegraphics[scale=0.16]{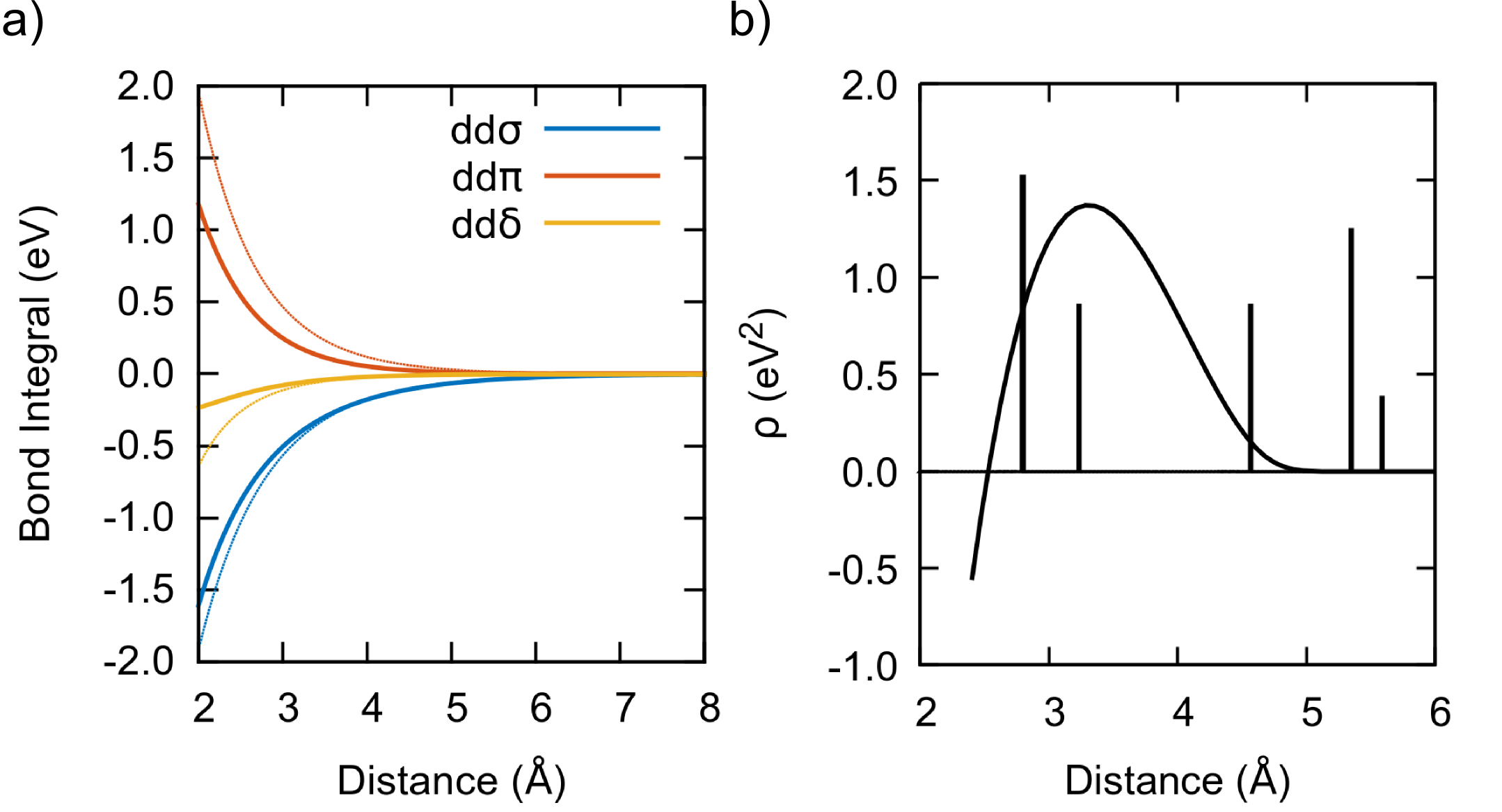}
\par\end{centering}
\caption{\textbf{a)} Distance dependence of the $dd\sigma$, $dd\pi$,
  and $dd\delta$ bond integrals before (thin lines) and after (thick
  lines) optimization. \textbf{b)} Distance dependence of the
  embedding function. The radial distribution function of bcc Ti at
  equilibrium volume (vertical bars) has been superimposed.}
\label{bonding}
\end{figure}

\begin{figure}
\begin{centering} 
\includegraphics[scale=0.2]{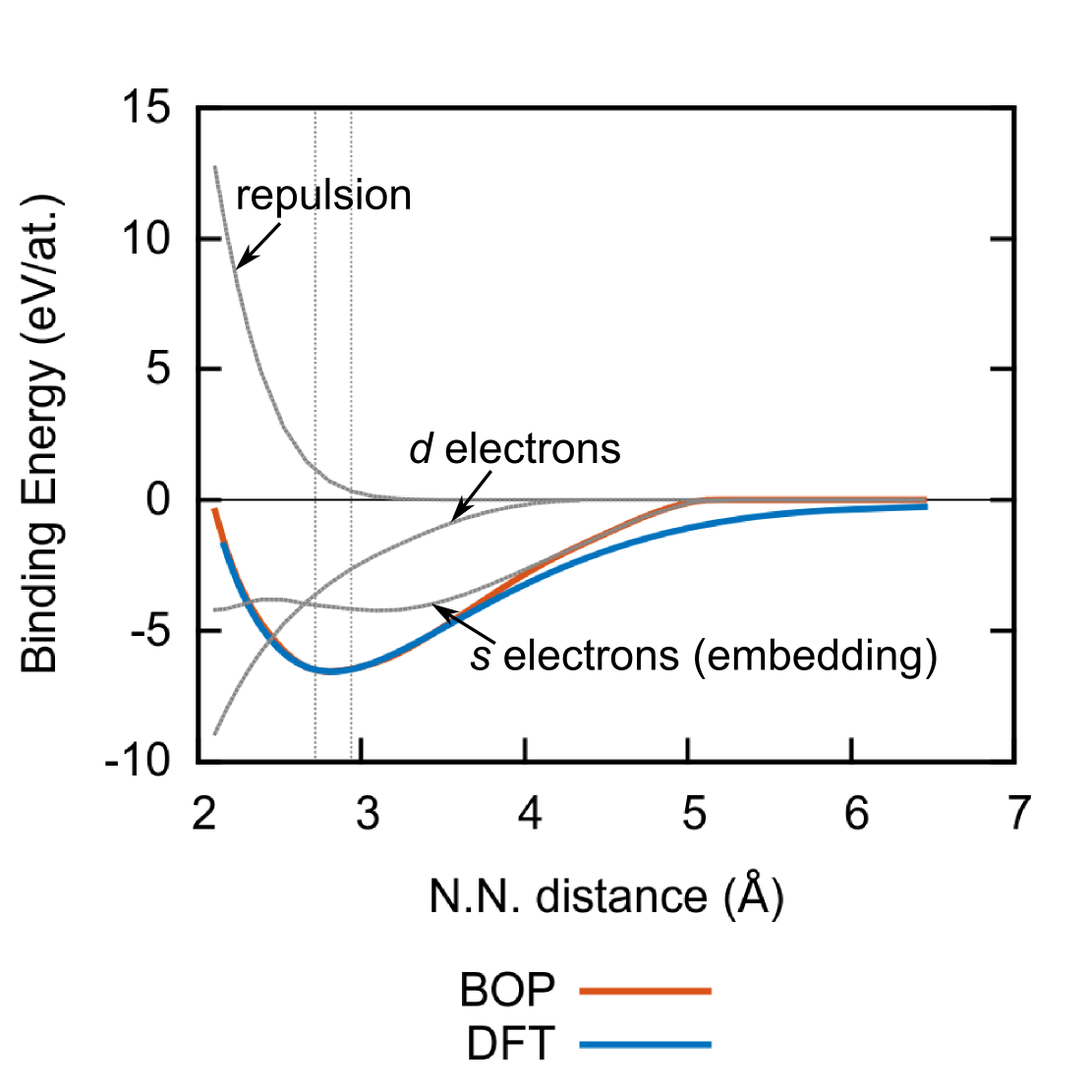}
\par\end{centering}
\caption{Binding energy of bcc Ti for BOP (red) and DFT (blue) as
  a function of the nearest neighbor distance. The BOP binding energy
  is decomposed into the bonding, repulsive, and embedding
  contributions (grey). The dashed vertical lines denote the volume
  range used in the fitting.}
\label{contrib}
\end{figure}

We optimized our interatomic potential using the Levenberg-Marquardt
\cite{Levenberg1944,Marquardt1963} least-squares method as implemented
in the Bond-Order Potential Characterization, Assessment, and Testing
(BOPcat) suite \cite{Ladines}.
The starting parameters for the bond integrals were taken
from a projection of the DFT wavefunction on an $sd$ orthogonal TB
basis set for the Ti dimer. The details of the projection method
can be found in Ref.~\onlinecite{Jenke2019}. In a
first step, we considered a full $sd$ model derived from the Ti
dimer and fitted the parameters $A$, $B$, and $C$ in the repulsive
term to reproduce the DFT binding energy-volume curves of the
$\omega$, hcp, fcc, bcc, and A15 structures as closely as possible. We
then substituted the explicit treatment of the $s$ electrons with the
embedding function and determined the four parameters of this function
to reproduce the same DFT data, keeping the bond integrals
and the repulsive term fixed. Finally, for the $\omega$, hcp, and bcc
structures only, we added to the fitting database the binding energies
of the elastically deformed structures and the forces resulting from
the small displacements method for the phonons. At this stage, the energy-volume curve of the dhcp phase was also included.
In the final optimization step, all 25 model parameters were adjusted.  The optimized values of
the parameters for our model are listed in Tab.~1 of the Supplemental
Material \cite{SuppMat}.

Fig.~\ref{bonding}a displays the distance dependence of the
$dd\sigma$, $dd\pi$, and $dd\delta$ bond integrals as derived from the
Ti dimer projections (thin lines) and after full optimization
(thick lines).  As expected, the range of the bond integrals becomes
shorter after optimization because in the bulk environment the
interactions are screened by the charge densities of neighboring atoms
\cite{Nguyen-Manh2000}.  It is worth 
noting that at a distance of 2.9~\AA, corresponding approximately to
the first nearest neighbor shell for most structures, the ratio of the optimized
bond integrals is
$dd\sigma:dd\pi:dd\delta=-60:31:-9$, that is close to the canonical ratio
employed in Sec.~\ref{canonical}.

Fig.~\ref{bonding}b shows the optimized embedding function
$\rho\left(r\right)$, superimposed on the radial distribution function of bcc
at the equilibrium volume. $\rho\left(r\right)$ has a maximum at approximately
3.3 \AA, roughly at the second nearest-neighbor, 
%in the equilibrium bcc structure, 
and decreases smoothly to zero for long interatomic distances. This
variation is consistent with our analysis of $sd$ and $d$ TB models obtained
by a projection of pseudo-atomic orbitals on DFT wave functions
\cite{Urban2011} and will be discussed in detail elsewhere
\cite{Mrovec:unpublished}.  At extremely short distances ($<$ 2.5 \AA),
outside the fitting range, $\rho\left(r\right)$ becomes negative. This is
interpreted as a many-body \textit{repulsive} overlap contribution due to
overlapping charge densities.

The contributions of the bonding, embedding, and repulsive terms for our BOP
for bcc Ti are marked by grey lines in Fig.~\ref{contrib}. The $d$ electrons
contribute approximately 40\% to the cohesive energy, while the remaining part is
due to the $s$ electrons. This agrees with the respective contributions to the
cohesive energy derived from $sd$ and $d$ orbital projections
\cite{Urban2011}.

The (total) binding energy calculated with our potential (red line in
Fig.~\ref{contrib}) agrees very well with the DFT binding energy (blue
line in Fig.~\ref{contrib}) even for interatomic distances well
outside the fitting range (delineated by the dashed vertical lines in
Fig.~\ref{contrib}).  This denotes an outstanding transferability of
our BOP to structures with both small and large densities. The BOP
binding energy starts deviating significantly from the DFT reference
only at large interatomic distances exceeding 4 \AA, where the
embedding function decreases to zero.

\section{Properties of bulk phases}
\label{simple_phases}

\begin{figure}
\begin{centering} 
\includegraphics[scale=0.15]{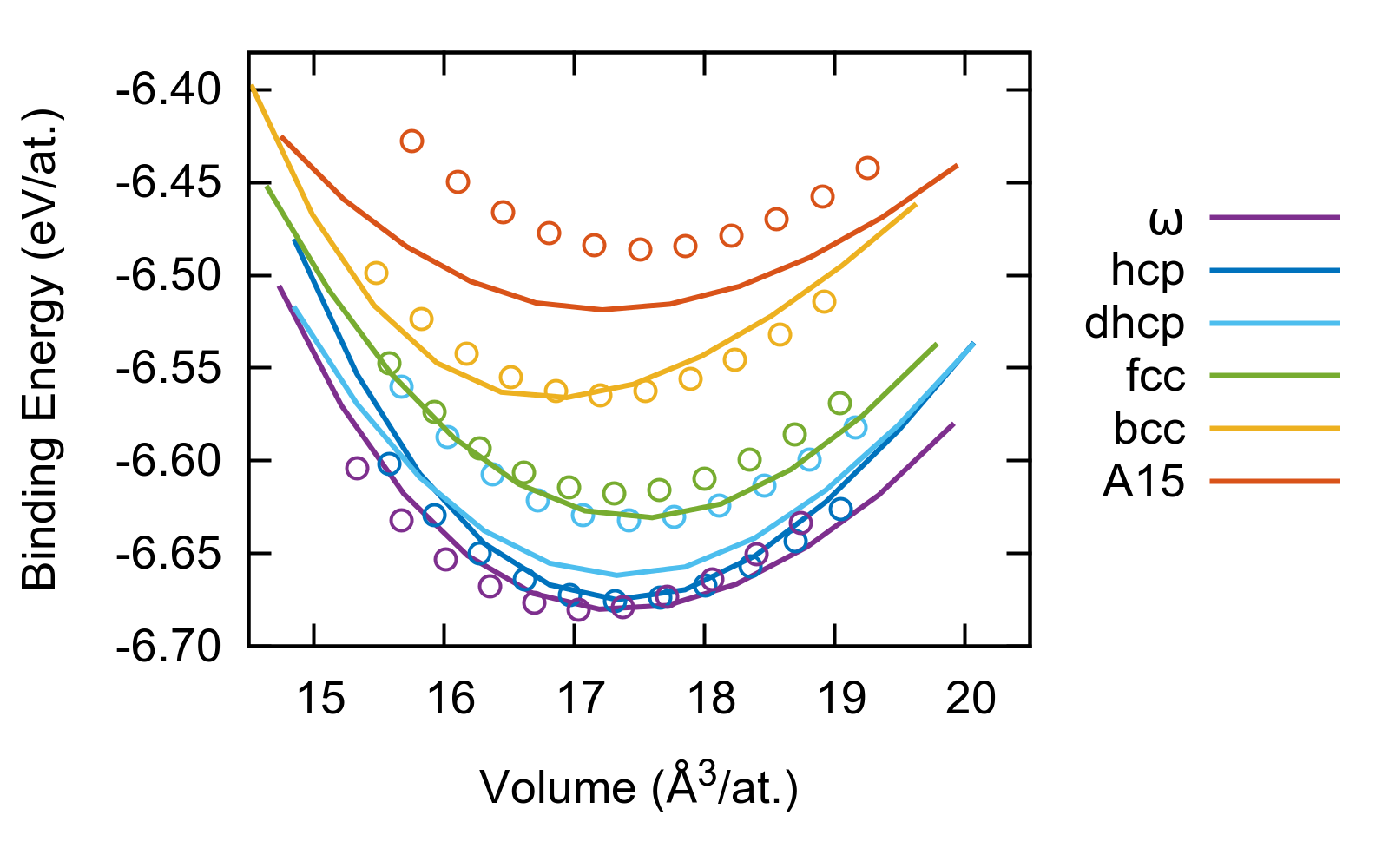}
\par\end{centering}
\caption{Binding energy-volume curves of the phases used in the fitting from the BOP (lines) and DFT (circles).}
\label{EV}
\end{figure}

\begin{figure*}
\begin{centering} 
\includegraphics[scale=0.07]{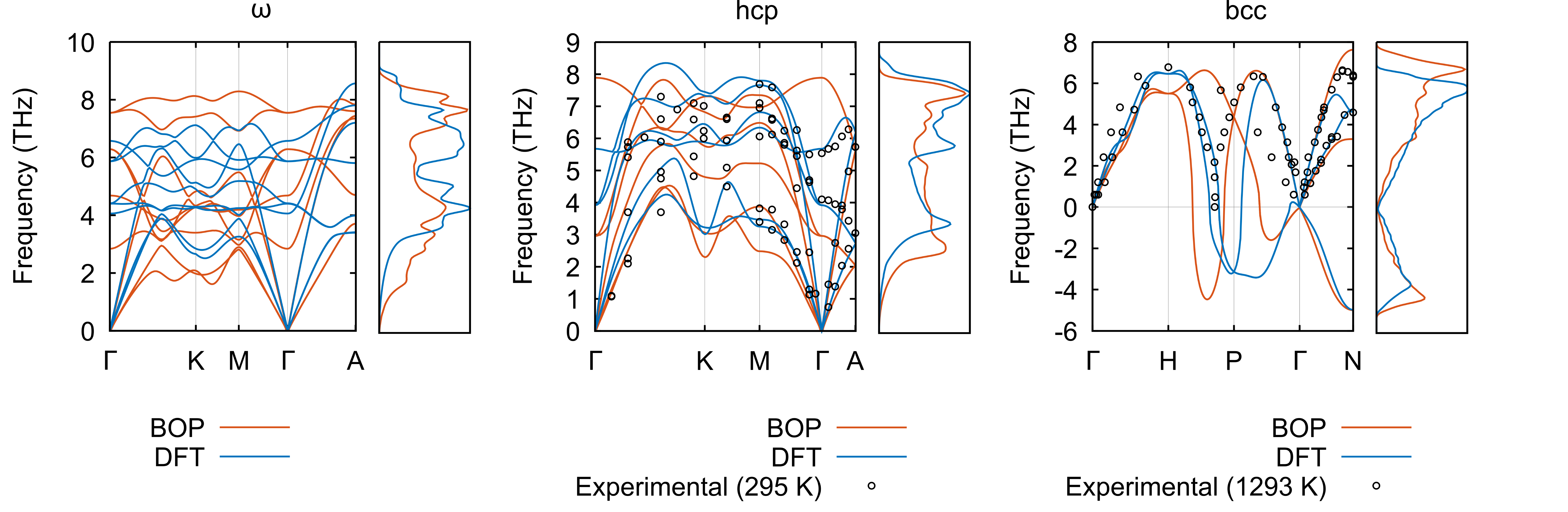}
\par\end{centering}
\caption{Phonon dispersion relations along high-symmetry paths in the
  Brillouin zone and phonon densities of modes of the $\omega$, hcp,
  and bcc phases. Experimental data taken from Refs.~\onlinecite{Stassis1979,Petry1991}}
\label{phonons}
\end{figure*}

The BOP (lines) and DFT (circles) binding energy-volume curves are
compared for the fitted structures in Fig.~\ref{EV}. 
The cohesive energies of
$\omega$, hcp, and bcc are very well captured by our potential, while the
energies of the fcc and A15 phases are slightly underestimated. The model also
underestimates the energy of the double-hcp (dhcp) phase, which 
%has a consequence on 
is important for
the properties of basal stacking faults (see Sec.~\ref{defects}).
In Tab.~\ref{prop} the equilibrium lattice parameters, bulk moduli, and
elastic constants of the $\omega$, hcp, and bcc phases as predicted by our
potential are compared  to our DFT results, the values computed with the BOP of Girshick \textit{et al.} \cite{Girshick1998}, the results of a
non-orthogonal TB model of Trinkle {\em et al.} \cite{Trinkle2006}, and to
available experimental values. For the
$\omega$ and hcp structures, the DFT lattice parameters are very well
reproduced by the BOP, while for bcc the lattice parameter is underestimated by
1\% with respect to DFT. This is reflected in an underestimation of the volume
of the bcc structure visible in Fig.~\ref{EV}. The quality of the bulk moduli
and elastic constants is almost as good as that of the non-orthogonal TB model
\cite{Trinkle2006}, and the deviation from the experimental measurements is only about twice as large as the deviation between DFT and experiments.
The present BOP describes the $\omega$ phase much better than the BOP by Girshick \textit{et al.}, who did not consider this phase in their parametrization, without compromising significantly the properties of the hcp and bcc structures.
It is also worth pointing out that the elastic constants
were not included directly in the fitting procedure of our potential, but they
are related to the curvature of the energy-deformation curves in our training
database.

\begin{table*}
\begin{ruledtabular}
\begin{tabular}{cccccc}
 &exp.&DFT (this work)&NOTB (Trinkle \textit{et al.}\cite{Trinkle2006})&BOP (Girshick \textit{et al.} \cite{Girshick1998})&BOP (this work)\tabularnewline
\hline
$\omega$\tabularnewline
$a$& 4.626 \cite{Akahama2001} & 4.563 & 4.580 & 4.520 & 4.575\tabularnewline
$c/a$& 0.608 \cite{Akahama2001} & 0.620 & 0.619 & 0.639 & 0.622\tabularnewline
$B$& 112 \cite{Tane2013} & 112 & -- & 81 & 106\tabularnewline
$C_{11}$& 179 \cite{Tane2013} & 193 & 184 & 146 & 151\tabularnewline
$C_{12}$& 90 \cite{Tane2013} & 79 & 90 & 79 & 89\tabularnewline
$C_{13}$& 61 \cite{Tane2013} & 52 & 52 & 40 & 60\tabularnewline
$C_{33}$& 228 \cite{Tane2013} & 241 & 261 & 118 & 234\tabularnewline
$C_{44}$& 71 \cite{Tane2013} & 56 & 100 & 12 & 28\tabularnewline
\hline
hcp\tabularnewline
$a$& 2.950 \cite{Fisher1964} & 2.928 & 2.940 & 2.954 & 2.922\tabularnewline
$c/a$& 1.587 \cite{Fisher1964} & 1.586 & 1.602 & 1.587 & 1.604\tabularnewline
$B$& 110 \cite{Fisher1964}& 123 & -- & 114 & 117\tabularnewline
$C_{11}$& 176 \cite{Fisher1964}& 196 & 155 & 176 & 170\tabularnewline
$C_{12}$& 87 \cite{Fisher1964} & 71 & 91 & 75 & 96\tabularnewline
$C_{13}$& 68 \cite{Fisher1964} & 83 & 79 & 84 & 86\tabularnewline
$C_{33}$& 191 \cite{Fisher1964}& 191 & 173 & 184 & 144\tabularnewline
$C_{44}$& 51 \cite{Fisher1964} & 39 & 65 & 51 & 29\tabularnewline
\hline
bcc\tabularnewline
$a$& 3.310 \cite{Petry1991} & 3.263 & 3.27 & 3.231 & 3.228\tabularnewline
$B$& 88 \cite{Ledbetter2004}, 118 \cite{Petry1991} & 105 & -- & 108 & 113\tabularnewline
$C_{11}$& 98 \cite{Ledbetter2004}, 134 \cite{Petry1991}& 104 & 87 & 95 & 83\tabularnewline
$C_{12}$& 83 \cite{Ledbetter2004}, 110 \cite{Petry1991} & 116 & 112 & 115 & 129\tabularnewline
$C_{44}$& 38 \cite{Ledbetter2004}, 36 \cite{Petry1991} & 36 & 31 & 58 & 37\tabularnewline
%$E_{f}^{(v1)}$& - & 2.92 \cite{Trinkle2006} & 2.85 & 3.29 & 1.27 \cite{Hashimoto1984} & 2.045 \cite{Ko2015} & 1.88 & 2.83 & - & - & - & -\tabularnewline
%$E_{f}^{(v2)}$& - & 1.57 \cite{Trinkle2006}& 1.34 & 1.66 & - & - & - & - & - & - & - & -\tabularnewline
\end{tabular}
\end{ruledtabular}
\caption{Structural properties of the $\omega$, hcp, and bcc phases. 
The experimental lattice parameters and elastic constants for the $\omega$ phase and the lattice parameters of hcp refer to ambient conditions.
The experimental elastic constants of hcp were measured at 4 K.  The lattice and elastic constants of bcc were measured at 1293 K for Ref.~\onlinecite{Petry1991} and at 1273 K for Ref.~\onlinecite{Ledbetter2004}. 
Lattice parameters are in \AA, and bulk moduli and elastic constants in GPa.
}%, and vacancy formation energies in eV. }
\label{prop}
\end{table*}

Fig.~\ref{phonons} presents the phonon dispersion relations and densities of
modes of the $\omega$, hcp, and bcc phases for our BOP, DFT, and, where
available, experiments. The acoustic phonons of the $\omega$ phase are
slightly underestimated with respect to DFT, while the optical phonons are
overestimated for all high-symmetry points except for the A zone boundary.
The underestimation of the acoustic branches is a consequence of the
underestimation of the $C_{11}$ elastic constant. These differences are also
reflected in the phonon density of modes. The spectrum of hcp matches DFT and
experiments quite closely, apart from softenings  at the K and A points and the
negative curvature of the optical branch at $\Gamma$, which is a feature common to most interatomic potentials for hcp
metals. The bcc phonons are also captured very  well by our BOP, and the 0~K
$\omega$- and $\alpha$-instabilities at 
$\frac{2}{3}\left[111\right]$ and $\left[110\right]$, respectively, are both present.

To check the performance of our interatomic potential in different atomic
environments, we computed the cohesive energy of some low-energy (up to 1 eV higher than the ground state) prototype structures using both DFT and BOP.
The results for the various prototypes, indicated by their
\textit{Strukturbericht} designations and names of the most common compounds
with that particular structure, are reported in Tab.~\ref{prototypes}. 
The structures included in the fit set are shown in bold. As
deduced by the relatively small differences between the BOP predictions and
the first principles data, the potential shows a remarkable transferability to
very different coordination polyhedra and even to exotic structures, rarely
considered during testing of interatomic potentials. The most significant
differences between the cohesive energies of DFT and BOP are the simple cubic
(sc), the A11 ($\alpha$-Ga), and A5 ($\beta$-Sn) structures, all characterized
by a relatively low cohesive energy. The average error for the considered
prototypes is 90 meV/at.

\begin{figure*}
\begin{centering} 
\includegraphics[scale=0.20]{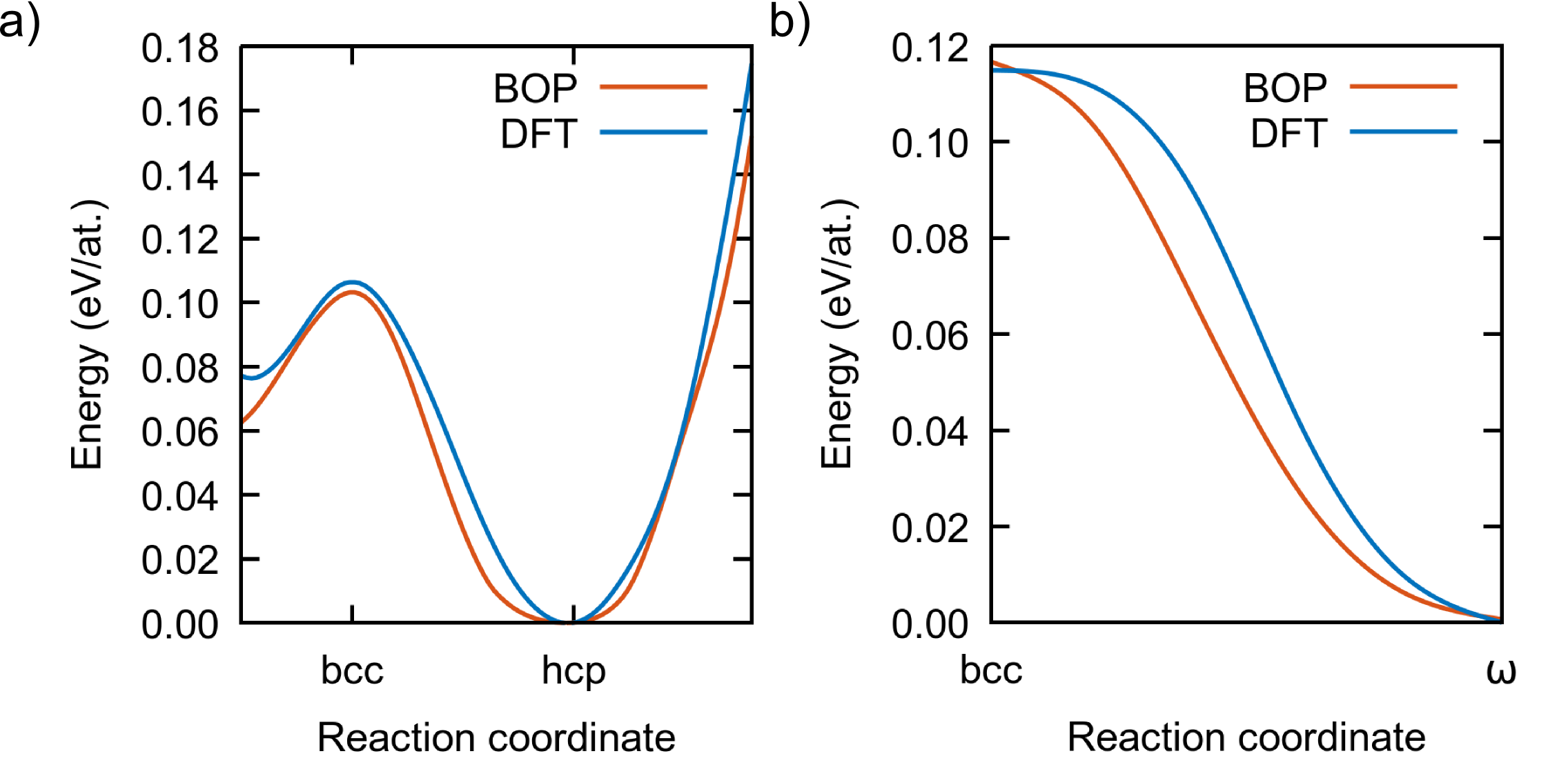}
\par\end{centering}
\caption{\textbf{a)} Hexagonal and \textbf{b)} bcc $\rightarrow \omega$ transformation paths calculated with the BOP and with DFT. The lowest energy for each path is set to zero.}
\label{tr_path}
\end{figure*}

To further test the BOP, we calculated the energy along the hexagonal and bcc $\rightarrow \omega$ transformation paths with DFT and with the present BOP, since these paths are crucial for the phase transitions in Ti.
The details of the hexagonal transformation path can be found in  Refs.~\onlinecite{Paidar1999,Luo2002,Cak2014};
the  bcc $\rightarrow \omega$ transition consists of a shuffling of pairs of atoms along the [111] direction of bcc, corresponding to the $\frac{2}{3}[111]$ phonon \cite{Ehemann2017}.
For simplicity, the volume of the unit cell was taken as constant along the transformations.
The results, displayed in Fig.~\ref{tr_path}, show that both paths are very well reproduced by our BOP, even if the intermediate points were not included during the fitting.

\begin{table*}
\begin{ruledtabular}
\begin{tabular}{cccc|cccc}
Prototype & $E_\text{coh}^{\text{(DFT)}}$  & $E_\text{coh}^{\text{(BOP)}}$  & $E_\text{coh}^{\text{(BOP)}}$ - $E_\text{coh}^{\text{(DFT)}}$ & Prototype & $E_\text{coh}^{\text{(DFT)}}$ & $E_\text{coh}^{\text{(BOP)}}$  & $E_\text{coh}^{\text{(BOP)}}$ - $E_\text{coh}^{\text{(DFT)}}$\tabularnewline
\hline 
\textbf{C32 ($\omega$)} & \textbf{6.680} & \textbf{6.680} & \textbf{0.000} & A17 (black P) & 6.517 & 6.563 & 0.046\tabularnewline
\textbf{A3 (hcp)} & \textbf{6.676} & \textbf{6.676} & \textbf{0.000} & A$_\text{b}$ ($\beta$-U) & 6.507 & 6.514 & 0.007\tabularnewline
$\omega_\text{def}$ \cite{Korbmacher} & 6.672 & 6.675 & 0.003 & \textbf{A15 (Cr$_{3}$Si)} & \textbf{6.486} & \textbf{6.519} & \textbf{0.033}\tabularnewline
C19 ($\alpha$-Sm) & 6.651 & 6.666 & 0.015 & A12 ($\alpha$-Mn) & 6.486 & 6.558 & 0.072\tabularnewline
\textbf{A3' (dhcp)} & 6.632 & 6.662 & 0.030 & A11 ($\alpha$-Ga) & 6.454 & 6.190 & -0.264\tabularnewline
\textbf{A1 (fcc)} & \textbf{6.617} & \textbf{6.631} & \textbf{0.014} & C14 (MgZn$_{2}$) & 6.426 & 6.392 & -0.034\tabularnewline
A14 (I$_{2}$) & 6.599 & 6.516 & -0.083 & C15 (Cu$_{2}$Mg) & 6.421 & 6.338 & -0.083\tabularnewline
A$_\text{c}$ ($\alpha$-Np) & 6.579 & 6.587 & 0.008 & A5 ($\beta$-Sn) & 6.243 & 6.541 & 0.298\tabularnewline
\textbf{A2 (bcc)} & \textbf{6.565} & \textbf{6.566} & \textbf{0.001} & A$_\text{h}$ (sc) & 5.838 & 6.345 & 0.507\tabularnewline
A13 ($\beta$-Mn) & 6.558 & 6.537 & -0.021 & A9 (graphite) & 5.681 & 5.405 & -0.276\tabularnewline
\end{tabular}
\end{ruledtabular}
\caption{Cohesive energies in eV of some low energy prototype structures with respect to a non-magnetic Ti atom with electronic configuration $\left[Ar\right]3s^{2}3d^{2}$.
The fitted structures are shown in bold. 
$\omega_\text{def}$ is a metastable defective $\omega$ structure, described recently by Korbmacher \textit{et al.} \cite{Korbmacher}, important in the bcc$\rightarrow \omega$ transformation in Ti.}
\label{prototypes}
\end{table*}

\section{Defect properties}
\label{defects}

%The properties of defects were completely absent from
%the fit set so it is both interesting and important to ana-
%lyze

The properties of defects were completely absent from the fit set and therefore 
constitute an important test for the predictive capabilities of the
developed potential for highly distorted atomic configurations around
fundamental defects.  We computed the formation energies of vacancies, low-index
surfaces and fundamental stacking faults in the hcp and $\omega$ phases.  The
obtained results are listed in Tab.~\ref{defprop} together with the results from
other TB models, the previous BOP, DFT, and experiments (where available).

The vacancy formation energies in both hcp and $\omega$ phases are
overestimated by our BOP, but the relative stability of vacancies in the
$\omega$ phase (which has two inequivalent Wyckoff positions and thus two
different sites for the vacancy) is well reproduced.  
The values of the surface formation energies
agree very well with the DFT values, in contrast to other classical potentials \cite{Ackland1992, Zhou2004, Hennig2008, Gibson2016} and the previous BOP \cite{Girshick1998, Girshick1998a}, although the relative ordering of the energetics does not correspond to DFT.
A relatively large systematic underestimation is obtained for the energies of the
fundamental stacking faults on the basal plane of hcp. All three calculated
stacking faults are about 100 mJ/m$^2$ lower than the reference DFT values.
This deviation is most likely related to underestimation of the dhcp energy
by our model. According to DFT, the dhcp structure is 44 meV/at.~less stable than
hcp, whereas our BOP predicts only 14 meV/at.  This discrepancy leads to the
underestimation of stacking fault energies. 
An even more severe underestimation of the stacking fault energies is observed in the BOP by Girshick \textit{et al.}, pointing out that the correct description of the energy difference between hcp and dhcp and thus of the stacking faults might be beyond the limitations of $d$-only models.
Even with the explicit inclusion of dhcp in the fitting database, the cohesive energy of this structure could not be improved without compromising the stability of the other phases.

\begin{table*}
\begin{ruledtabular}
\begin{tabular}{lccccc}
 &exp.&DFT (this work)&NOTB (Trinkle \textit{et al.}\cite{Trinkle2006})&BOP (Girshick \textit{et al.} \cite{Girshick1998})&BOP (this work)\tabularnewline
\hline 
hcp defects& & & & & \tabularnewline
$E_\text{vac}^\text{f}$ [eV] & $> 1.70$ \cite{DeBoer1988} & 1.92-2.07 \cite{Raji2009} & 1.81 & 2.33 & 2.80 \tabularnewline

$E_\text{surf}(0001)$        [mJ/m$^2$] & 2100 \cite{Koeppers1997} &  1939 \cite{Hennig2008} & -- & 1454 & 2083 \tabularnewline
$E_\text{surf}(1\bar{1}00)$  [mJ/m$^2$] & 1920 \cite{Tyson1977}  &  2451 \cite{Hennig2008} & -- & 1571 & 2337 \tabularnewline
$E_\text{surf}(11\bar{2}0)$  [mJ/m$^2$] & --                     &  1875 \cite{Hennig2008} & -- & 1741 & 2271 \tabularnewline

$E_\text{sf}(ISF1)$  [mJ/m$^2$] & -- &  149 \cite{Benoit2012} & -- & 38 & 62 \tabularnewline
$E_\text{sf}(ISF2)$  [mJ/m$^2$] & -- &  259 \cite{Benoit2012} & -- & 106 & 160 \tabularnewline
$E_\text{sf}(ESF)$   [mJ/m$^2$] & -- &  353 \cite{Benoit2012} & -- & 171 & 256 \tabularnewline
\hline
$\omega$ defects & & & & & \tabularnewline
$E_\text{vac}^\text{f}(A)$ [eV] & -- & 2.92 \cite{Trinkle2006} & 2.85 & 2.78 & 3.34 \tabularnewline
$E_\text{vac}^\text{f}(B)$ [eV] & -- & 1.57 \cite{Trinkle2006} & 1.34 & 0.68 & 1.61 \tabularnewline

$E_\text{surf}(0001)$        [mJ/m$^2$] & -- &  2131 \cite{Hennig2008} & -- & 1764 & 2527 \tabularnewline
$E_\text{surf}(1\bar{1}00)$  [mJ/m$^2$] & -- &  2179 \cite{Hennig2008} & -- & 1776 & 2490 \tabularnewline
$E_\text{surf}(11\bar{2}0)$  [mJ/m$^2$] & -- &  2435 \cite{Hennig2008} & -- & 1460 & 2099 \tabularnewline

\end{tabular}
\end{ruledtabular}
\caption{Formation energies of vacancies, surfaces and
  stacking faults in the hcp and $\omega$ phases.  The NOTB data are
  taken from Ref.~\onlinecite{Trinkle2006}.
  $E_\text{surf}(0001)$ for the $\omega$ phase refers to the surface termination with the high-density plane. }
\label{defprop}
\end{table*}

\section{Thermodynamic properties}
\label{temperature}

\subsection{Temperature-induced phase transformations}

\begin{figure}
\begin{centering} 
\includegraphics[scale=0.23]{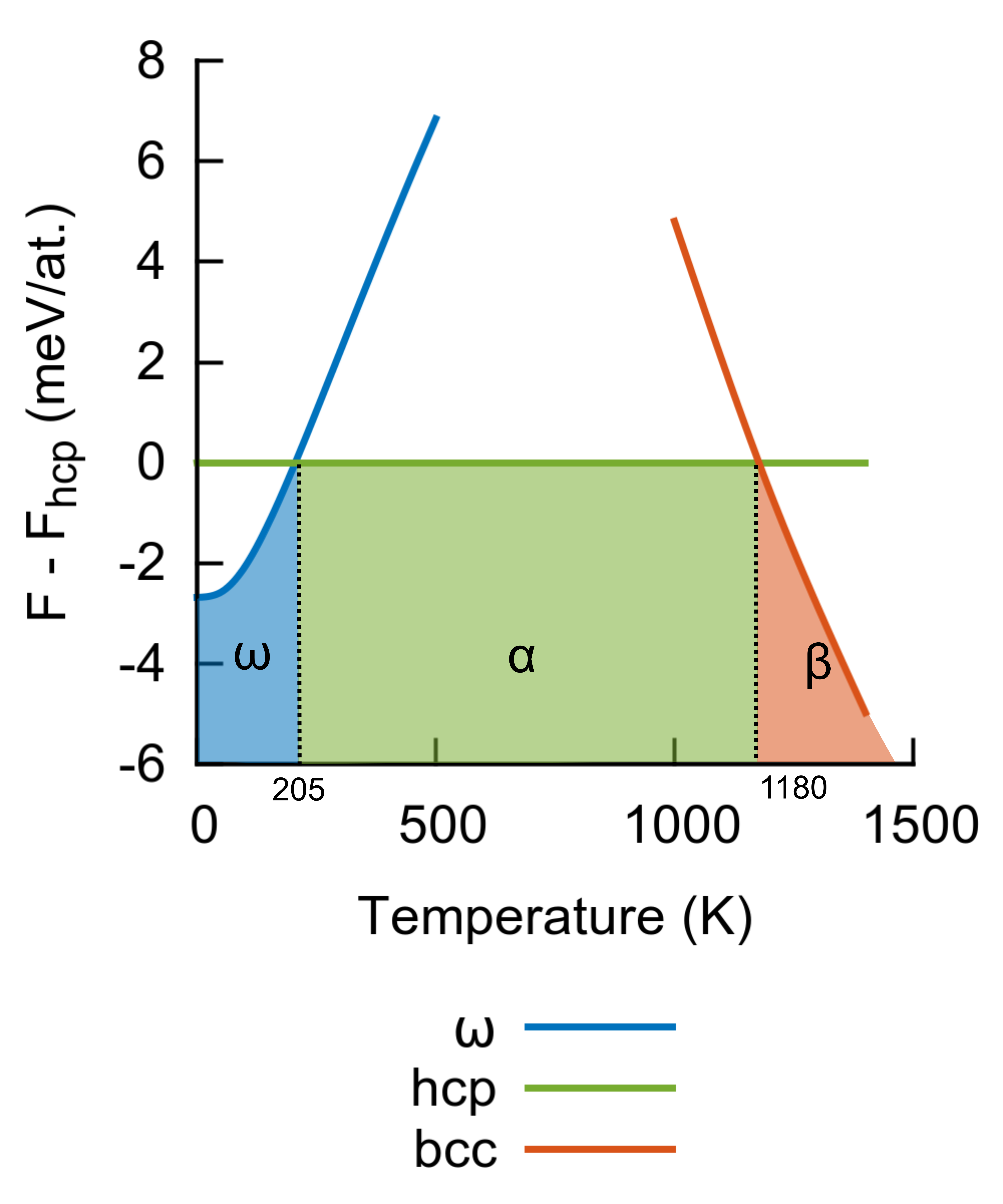}
\par\end{centering}
\caption{Helmholtz free energy differences with respect to the hcp phase as a function of temperature. At zero pressure the phase with the lowest free energy is the most stable phase.}
\label{free_energy}
\end{figure}

Ti exhibits a complex phase diagram that is challenging to
reproduce using empirical interatomic potentials.  We employed our
BOP in molecular dynamics (MD) simulations at finite
temperatures and focused on the phase transformations $\omega\rightarrow$hcp,
hcp$\rightarrow$bcc, and bcc$\rightarrow$liquid at zero pressure.  To estimate
the transition temperatures for the two martensitic transformations
$\omega\rightarrow$hcp and hcp$\rightarrow$bcc we computed the free energies
$F\left(T,V\right)$ of the $\omega$, hcp, and bcc phases. For $\omega$ and
hcp, since these phases are dynamically stable at 0 K, we employed the
harmonically assisted temperature integration method detailed in
Ref.~\onlinecite{Moustafa2017}.  The free energy can then be decomposed as the sum
of a harmonic term, which depends on the phonon density of modes
$g_\text{ph}\left(\omega,V\right)$ at a given volume $V$, and an anharmonic
contribution $F_\text{ah}$,
\[
F\left(T,V\right)=k_\text{B}T\int_{0}^{\omega_\text{max}}g_\text{ph}\left(\omega,V\right)\ln\left[2\sinh\left(\frac{\hbar\omega}{2k_\text{B}T}\right)\right]d\omega
\]
\begin{equation}
+F_\text{ah}\left(T,V\right) \quad .
\end{equation}
Following Ref.~\onlinecite{Moustafa2017}, we computed $F_\text{ah}\left(T,V\right)$ as
\begin{equation}
F_\text{ah}\left(T,V\right)=-T\cdot\int_{0}^{T}\frac{1}{\hat{T}^{2}}\left\langle U_\text{pot}-U_\text{pot}^{(0)}+\frac{1}{2}\vec{F}\cdot\Delta\vec{r}\right\rangle _{\hat{T},V}d\hat{T} \quad ,
\label{anh}
\end{equation}
where $U_\text{pot}$, $\vec{F}$, and $\Delta\vec{r}$ are respectively the potential
energy, forces and displacements from the equilibrium positions extracted from
MD simulations at various temperatures and volumes, and $U_\text{pot}^{(0)}$ is the
energy of the equilibrium $\omega$ or hcp phases.  The thermal averages in
Eq.~\eqref{anh} were calculated from MD trajectories in the $NVT$ ensemble
with a duration of 10 ps after complete equilibration with a Langevin
thermostat for 5 different volumes.  For the $\omega$ phase we employed a
$4\times4\times6$ supercell while for the hcp phase a $6\times6\times4$
supercell, with a total of 288 atoms for both structures. For each
temperature, the obtained free energy-volume curves were fitted using the
Birch-Murnaghan equation \cite{Murnaghan1944,Birch1947} to determine the value
of the zero-pressure (Helmholtz) free energy.

Since the bcc structure is not stable at 0 K, temperature integration 
 as in Eq.~\eqref{anh} is not possible. To calculate the free
energy of the bcc phase, we instead employed the standard Frenkel-Ladd method
\cite{Frenkel1984} to integrate the free energy difference between our
potential $U_{1}$ and a reference potential $U_{0}$,
\begin{equation}
\Delta F=\int_{0}^{1}\left\langle U_{1}-U_{0}\right\rangle _{\lambda}d\lambda \quad ,
\label{Frenkel_Ladd}
\end{equation}
choosing as the reference system an Einstein crystal with potential energy
\begin{equation}
U_{0}=\sum_{i=1}^{N}\frac{1}{2}k\Delta r_{i}^{2} \quad ,
\end{equation}
with $k=5$ eV/\AA$^{2}$.
The thermal averages $\left\langle U_{1}-U_{0}\right\rangle _{\lambda}$ were
again calculated in the $NVT$ ensemble for 10 ps using a $6\times6\times6$ bcc
cubic supercell with 432 atoms. The volume was varied for each
temperature so that the total pressure was zero.  The integral in
Eq.~\eqref{Frenkel_Ladd} was evaluated using 15 values of the switching
parameter $\lambda$.

Fig.~\ref{free_energy} presents the Helmholtz free energy differences between
$\omega$ and hcp and between bcc and hcp as a function of temperature. 
The energy difference between $\omega$ and hcp at 0 K reduces to 3 meV/at.~if the
zero point energy is considered. Our BOP predicts a phase transition between
$\omega$ and hcp at 205 K, in good agreement with non-orthogonal tight-binding
(280 K) \cite{Rudin2004}. This transition has never been measured
experimentally at zero pressure because of the large free energy barrier that
separates the two phases; however, the transformation temperature must be
below room temperature, as correctly predicted by our BOP. The phase
transition between hcp and bcc occurs for our potential at 1180 K, in
excellent agreement with experiments that detect the transition at 1155 K
\cite{Holland1963}.

Finally, we also estimated the melting point of Ti with our interatomic
potential by gradually heating the bcc phase until melting was
observed. For this calculation a $15\times9\times9$ bcc supercell in a slab
geometry with two free surfaces was employed.  
The $\{100\}$ free surfaces in
the periodic cells were separated in  $x$ direction by roughly 5 nm of
vacuum. The dimensions along the [010] and [001] directions were adjusted for each temperature to minimize the stresses.
By analyzing the radial distribution function of the slab,
we estimated a melting temperature of $2000\pm200$ K, in good
agreement with the experimental melting point (1941 K).

In general, the predictions of our BOP model for finite temperature
thermodynamic properties show an impressive agreement with experiments even
though the fitting database was composed only of 0 K data. This is perhaps the
best exemplification of the excellent transferability of our potential to
properties not included in the training set.

\subsection{Pressure-induced phase transformations}

\begin{figure}
\begin{centering} 
\includegraphics[scale=0.23]{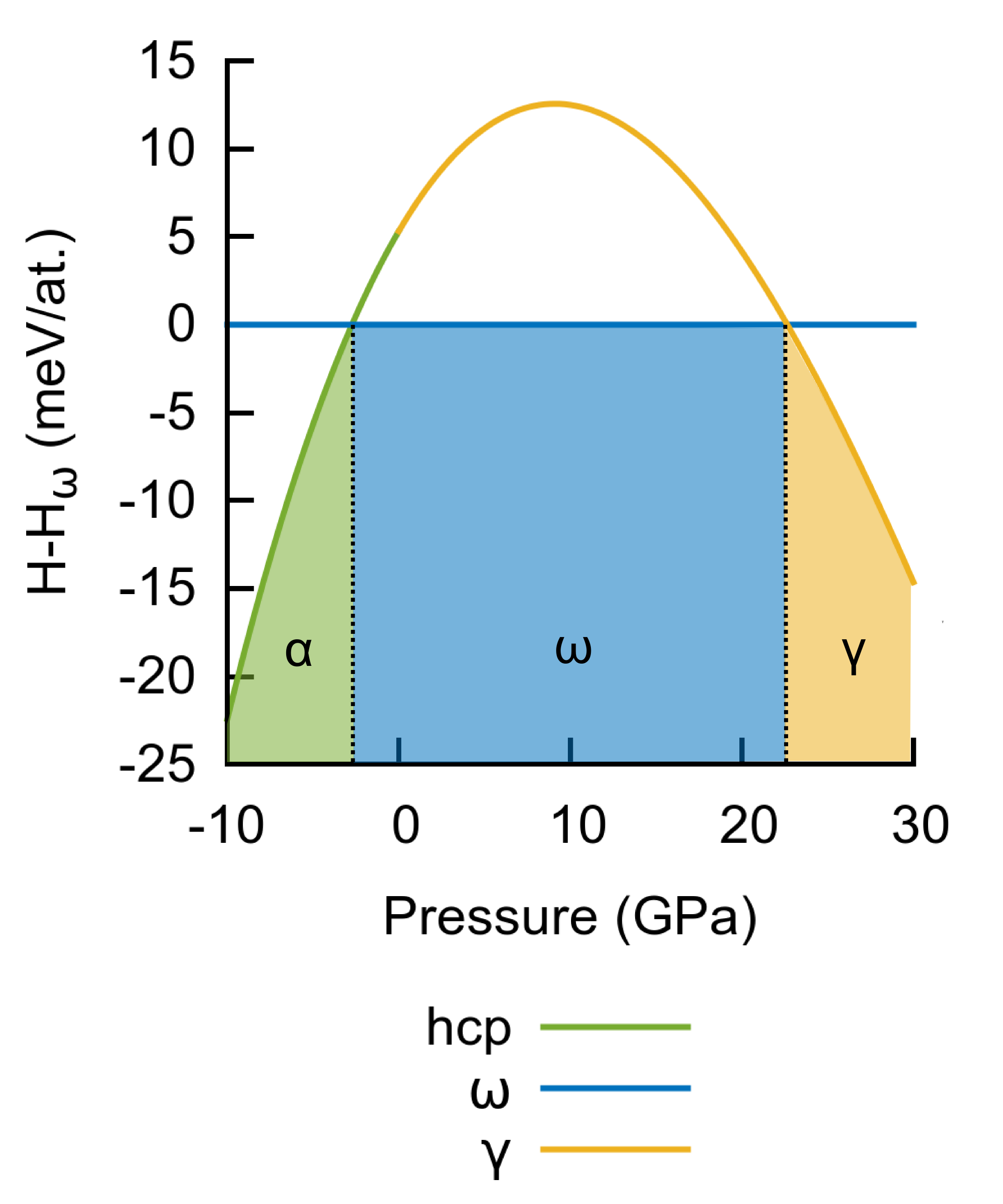}
\par\end{centering}
\caption{Enthalpy differences with respect to the $\omega$ phase as a function
  of pressure. At zero temperature the phase with the lowest enthalpy is the
  most stable phase.}
\label{enthalpy}
\end{figure}

We also analyzed the predictions of our potential at non-zero pressure by
varying the volume of the unit cell of the $\omega$, hcp, and $\gamma$ phases.
At each volume, we optimized the $c/a$ ratio of the $\omega$ and hcp phases,
and the $b/a$ and $c/a$ ratios and the atomic positions of the $\gamma$ phase.
The obtained energy-volume data were then fitted using the
Birch-Murnaghan equation of state \cite{Murnaghan1944,Birch1947}.  The
pressure and the enthalpy were evaluated according to 
\begin{equation}
P=-\frac{\partial E}{\partial V} \rightarrow H=E+PV  \quad . 
\end{equation}
Fig.~\ref{enthalpy} illustrates the enthalpy difference between hcp and
$\omega$ and between $\gamma$ and $\omega$ as a function of pressure. At
negative pressure (expanded volume), the hcp phase becomes more stable than
$\omega$ at -3 GPa, which agrees well with the DFT value of -5 GPa
\cite{Hennig2008}. At high pressures, the $\gamma$ phase is stabilized, but the
transition pressure of 23 GPa predicted by BOP is too low compared to
experimental data ($116-128$~GPa \cite{Vohra2001,Akahama2001}).  This
is, however, not unexpected as this phase transformation is known to be due to
an $s$-$d$ transition \cite{Vohra2001}: at high pressure, the long-ranged $s$
orbitals become unfavourable and a fraction of $s$-electrons is promoted
to $d$ orbitals.  This in turn increases the $d$-band filling and
stabilizes orthorhombic structures with respect to hexagonal ones.  The form
of our embedding function, which mimics the contribution from the $s$
electrons, is clearly too simple to  quantitatively capture this mechanism.
Nevertheless, the BOP model can reproduce qualitatively the correct sequence
hcp$\rightarrow \omega \rightarrow \gamma$ with increasing pressure.

\section{Conclusions}
\label{sec:conclusions}

We developed a bond-order potential for Ti that
retains the essential features of the electronic structure of this
element without sacrificing the computational efficiency, thanks to
the linear scaling analytical expressions for the energy and forces.
The small number of parameters in our model did not preclude an accuracy
comparable to more complex parametrizations regarding the structural
properties of the $\omega$, hcp, and bcc phases. On the contrary,
the choice of a very simple, physically motivated model lead to
an extraordinary transferability to various atomic configurations not considered in the fitting procedure, including diverse
structures and prototypes not tested before. This transferability
is also reflected in a good reproducibility  of the energetics of some extended
defects, and very accurate thermodynamic properties.
The limitations of this potential include the mechanisms that involve critically the $s$-electrons, such as the pressure-induced $\omega \rightarrow \gamma$ transition, and the stacking fault energies.
Nevertheless, within its clearly delineated range of applicability, we
believe that our potential is suitable not only for the atomistic
characterization of the stable phases of Ti but also
to explore new  mechanisms in this intriguing material.

%\section*{Acknowledgements}
\begin{acknowledgements}
A.F., J.R. and R.D.~acknowledge financial support from the Deutsche Forschungsgemeinschaft (DFG) within the research unit FOR 1766 (High Temperature Shape Memory Alloys, http://www.for1766.de, project number 200999873, sub-group TP3), and Y.L.~and R.D.~through projects 405621217 and 405602047.
M.S.~acknowledges funding from the International Max Planck Research School (IMPRS) SurMat.
Part of the calculations presented in this work were performed on the Tetralith and Sigma clusters of the Swedish National Infrastructure for Computing (SNIC) at the National Supercomputer Centre (NSC) in Link\"oping, and on the Beskow cluster at the Center for High Performance Computing (PDC) in Stockholm.
The authors are grateful to Thomas Hammerschmidt and Alvin Ladines for the support with the BOPfox and BOPcat programs, respectively.
A.F.~thanks Tony Paxton for helpful suggestions on the fitting strategy and Sergej Starikov for fruitful discussions on the free energy of bcc Ti and the melting point.
\end{acknowledgements}

\bibliography{./bib/biblio}

\begin{thebibliography}{102}
\expandafter\ifx\csname natexlab\endcsname\relax\def\natexlab#1{#1}\fi
\expandafter\ifx\csname bibnamefont\endcsname\relax
  \def\bibnamefont#1{#1}\fi
\expandafter\ifx\csname bibfnamefont\endcsname\relax
  \def\bibfnamefont#1{#1}\fi
\expandafter\ifx\csname citenamefont\endcsname\relax
  \def\citenamefont#1{#1}\fi
\expandafter\ifx\csname url\endcsname\relax
  \def\url#1{\texttt{#1}}\fi
\expandafter\ifx\csname urlprefix\endcsname\relax\def\urlprefix{URL }\fi
\providecommand{\bibinfo}[2]{#2}
\providecommand{\eprint}[2][]{\url{#2}}

\bibitem[{\citenamefont{Ashby and Jones}(1998)}]{Ashby1998}
\bibinfo{author}{\bibfnamefont{M.~F.} \bibnamefont{Ashby}} \bibnamefont{and}
  \bibinfo{author}{\bibfnamefont{D.~R.~H.} \bibnamefont{Jones}},
  \emph{\bibinfo{title}{Engineering Materials 2}}
  (\bibinfo{publisher}{Elsevier}, \bibinfo{address}{Oxford},
  \bibinfo{year}{1998}).

\bibitem[{\citenamefont{Buehler et~al.}(1963)\citenamefont{Buehler, Gilfrich,
  and Wiley}}]{Buehler1963}
\bibinfo{author}{\bibfnamefont{W.~J.} \bibnamefont{Buehler}},
  \bibinfo{author}{\bibfnamefont{J.~V.} \bibnamefont{Gilfrich}},
  \bibnamefont{and} \bibinfo{author}{\bibfnamefont{R.~C.} \bibnamefont{Wiley}},
  \bibinfo{journal}{J. Appl. Phys.} \textbf{\bibinfo{volume}{34}},
  \bibinfo{pages}{1475} (\bibinfo{year}{1963}).

\bibitem[{\citenamefont{Doonkersloot and Vucht}(1970)}]{Doonkersloot1970}
\bibinfo{author}{\bibfnamefont{H.~C.} \bibnamefont{Doonkersloot}}
  \bibnamefont{and} \bibinfo{author}{\bibfnamefont{V.}~\bibnamefont{Vucht}},
  \bibinfo{journal}{J Less Common Met.} \textbf{\bibinfo{volume}{20}},
  \bibinfo{pages}{83} (\bibinfo{year}{1970}).

\bibitem[{\citenamefont{Kim et~al.}(2006)\citenamefont{Kim, Ikehara, Kim,
  Hosoda, and Miyazaki}}]{Kim2006}
\bibinfo{author}{\bibfnamefont{H.~Y.} \bibnamefont{Kim}},
  \bibinfo{author}{\bibfnamefont{Y.}~\bibnamefont{Ikehara}},
  \bibinfo{author}{\bibfnamefont{J.~I.} \bibnamefont{Kim}},
  \bibinfo{author}{\bibfnamefont{H.}~\bibnamefont{Hosoda}}, \bibnamefont{and}
  \bibinfo{author}{\bibfnamefont{S.}~\bibnamefont{Miyazaki}},
  \bibinfo{journal}{Acta Mater.} \textbf{\bibinfo{volume}{54}},
  \bibinfo{pages}{2419} (\bibinfo{year}{2006}).

\bibitem[{\citenamefont{Bagarjatskii et~al.}(1958)\citenamefont{Bagarjatskii,
  Nosova, and Tagunova}}]{Bagaryatskii1958}
\bibinfo{author}{\bibfnamefont{Y.~A.} \bibnamefont{Bagarjatskii}},
  \bibinfo{author}{\bibfnamefont{G.~I.} \bibnamefont{Nosova}},
  \bibnamefont{and} \bibinfo{author}{\bibfnamefont{T.~V.}
  \bibnamefont{Tagunova}}, \bibinfo{journal}{Dokl. Akad. Nauk SSSR}
  \textbf{\bibinfo{volume}{122}}, \bibinfo{pages}{593} (\bibinfo{year}{1958}).

\bibitem[{\citenamefont{Ferrari et~al.}(2019)\citenamefont{Ferrari,
  Sangiovanni, Rogal, and Drautz}}]{Ferrari2019}
\bibinfo{author}{\bibfnamefont{A.}~\bibnamefont{Ferrari}},
  \bibinfo{author}{\bibfnamefont{D.~G.} \bibnamefont{Sangiovanni}},
  \bibinfo{author}{\bibfnamefont{J.}~\bibnamefont{Rogal}}, \bibnamefont{and}
  \bibinfo{author}{\bibfnamefont{R.}~\bibnamefont{Drautz}},
  \bibinfo{journal}{Phys. Rev. B} \textbf{\bibinfo{volume}{99}},
  \bibinfo{pages}{094107} (\bibinfo{year}{2019}).

\bibitem[{\citenamefont{Endoh et~al.}(2017)\citenamefont{Endoh, Tahara,
  Inamura, and Hosoda}}]{Endoh2017}
\bibinfo{author}{\bibfnamefont{K.}~\bibnamefont{Endoh}},
  \bibinfo{author}{\bibfnamefont{M.}~\bibnamefont{Tahara}},
  \bibinfo{author}{\bibfnamefont{T.}~\bibnamefont{Inamura}}, \bibnamefont{and}
  \bibinfo{author}{\bibfnamefont{H.}~\bibnamefont{Hosoda}},
  \bibinfo{journal}{J. Alloys Compd.} \textbf{\bibinfo{volume}{695}},
  \bibinfo{pages}{76} (\bibinfo{year}{2017}).

\bibitem[{\citenamefont{Saito et~al.}(2003)\citenamefont{Saito, Furuta, Hwang,
  Kuramoto, Nishino, Suzuki, Chen, Yamada, Ito, Seno et~al.}}]{Saito2003}
\bibinfo{author}{\bibfnamefont{T.}~\bibnamefont{Saito}},
  \bibinfo{author}{\bibfnamefont{T.}~\bibnamefont{Furuta}},
  \bibinfo{author}{\bibfnamefont{J.-H.} \bibnamefont{Hwang}},
  \bibinfo{author}{\bibfnamefont{S.}~\bibnamefont{Kuramoto}},
  \bibinfo{author}{\bibfnamefont{K.}~\bibnamefont{Nishino}},
  \bibinfo{author}{\bibfnamefont{N.}~\bibnamefont{Suzuki}},
  \bibinfo{author}{\bibfnamefont{R.}~\bibnamefont{Chen}},
  \bibinfo{author}{\bibfnamefont{A.}~\bibnamefont{Yamada}},
  \bibinfo{author}{\bibfnamefont{K.}~\bibnamefont{Ito}},
  \bibinfo{author}{\bibfnamefont{Y.}~\bibnamefont{Seno}}, \bibnamefont{et~al.},
  \bibinfo{journal}{Science} \textbf{\bibinfo{volume}{300}},
  \bibinfo{pages}{464} (\bibinfo{year}{2003}).

\bibitem[{\citenamefont{Vohra and Spencer}(2001)}]{Vohra2001}
\bibinfo{author}{\bibfnamefont{Y.~G.} \bibnamefont{Vohra}} \bibnamefont{and}
  \bibinfo{author}{\bibfnamefont{P.~T.} \bibnamefont{Spencer}},
  \bibinfo{journal}{Phys. Rev. Lett.} \textbf{\bibinfo{volume}{86}},
  \bibinfo{pages}{3068} (\bibinfo{year}{2001}).

\bibitem[{\citenamefont{Akahama et~al.}(2001)\citenamefont{Akahama, Kawamura,
  and Le~Bihan}}]{Akahama2001}
\bibinfo{author}{\bibfnamefont{Y.}~\bibnamefont{Akahama}},
  \bibinfo{author}{\bibfnamefont{H.}~\bibnamefont{Kawamura}}, \bibnamefont{and}
  \bibinfo{author}{\bibfnamefont{T.}~\bibnamefont{Le~Bihan}},
  \bibinfo{journal}{Phys. Rev. Lett.} \textbf{\bibinfo{volume}{87}},
  \bibinfo{pages}{275503} (\bibinfo{year}{2001}).

\bibitem[{\citenamefont{Trinkle}(2003)}]{Trinkle2003}
\bibinfo{author}{\bibfnamefont{D.~R.} \bibnamefont{Trinkle}}, Ph.D. thesis,
  \bibinfo{school}{Ohio State University} (\bibinfo{year}{2003}).

\bibitem[{\citenamefont{Daw and Baskes}(1984)}]{Daw1984}
\bibinfo{author}{\bibfnamefont{M.~S.} \bibnamefont{Daw}} \bibnamefont{and}
  \bibinfo{author}{\bibfnamefont{M.~I.} \bibnamefont{Baskes}},
  \bibinfo{journal}{Phys. Rev. B} \textbf{\bibinfo{volume}{29}},
  \bibinfo{pages}{6443} (\bibinfo{year}{1984}).

\bibitem[{\citenamefont{Baskes}(1992)}]{Baskes1992}
\bibinfo{author}{\bibfnamefont{M.~I.} \bibnamefont{Baskes}},
  \bibinfo{journal}{Phys. Rev. B} \textbf{\bibinfo{volume}{46}},
  \bibinfo{pages}{2727} (\bibinfo{year}{1992}).

\bibitem[{\citenamefont{Ackland}(1992)}]{Ackland1992}
\bibinfo{author}{\bibfnamefont{G.~J.} \bibnamefont{Ackland}},
  \bibinfo{journal}{Phylos. Mag. A} \textbf{\bibinfo{volume}{66}},
  \bibinfo{pages}{917} (\bibinfo{year}{1992}).

\bibitem[{\citenamefont{Zope and Mishin}(2003)}]{Zope2003}
\bibinfo{author}{\bibfnamefont{R.~R.} \bibnamefont{Zope}} \bibnamefont{and}
  \bibinfo{author}{\bibfnamefont{Y.}~\bibnamefont{Mishin}},
  \bibinfo{journal}{Phys. Rev. B} \textbf{\bibinfo{volume}{68}},
  \bibinfo{pages}{024102} (\bibinfo{year}{2003}).

\bibitem[{\citenamefont{Zhou et~al.}(2004)\citenamefont{Zhou, Johnson, and
  Wadley}}]{Zhou2004}
\bibinfo{author}{\bibfnamefont{X.}~\bibnamefont{Zhou}},
  \bibinfo{author}{\bibfnamefont{R.}~\bibnamefont{Johnson}}, \bibnamefont{and}
  \bibinfo{author}{\bibfnamefont{H.}~\bibnamefont{Wadley}},
  \bibinfo{journal}{Phys. Rev. B} \textbf{\bibinfo{volume}{69}},
  \bibinfo{pages}{144113} (\bibinfo{year}{2004}).

\bibitem[{\citenamefont{Ko et~al.}(2015)\citenamefont{Ko, Grabowski, and
  Neugebauer}}]{Ko2015}
\bibinfo{author}{\bibfnamefont{W.-S.} \bibnamefont{Ko}},
  \bibinfo{author}{\bibfnamefont{B.}~\bibnamefont{Grabowski}},
  \bibnamefont{and}
  \bibinfo{author}{\bibfnamefont{J.}~\bibnamefont{Neugebauer}},
  \bibinfo{journal}{Phys. Rev. B} \textbf{\bibinfo{volume}{92}},
  \bibinfo{pages}{134107} (\bibinfo{year}{2015}).

\bibitem[{\citenamefont{Gibson et~al.}(2016)\citenamefont{Gibson, Srinivasan,
  Baskes, Miller, and Wilson}}]{Gibson2016}
\bibinfo{author}{\bibfnamefont{J.}~\bibnamefont{Gibson}},
  \bibinfo{author}{\bibfnamefont{S.}~\bibnamefont{Srinivasan}},
  \bibinfo{author}{\bibfnamefont{M.}~\bibnamefont{Baskes}},
  \bibinfo{author}{\bibfnamefont{R.}~\bibnamefont{Miller}}, \bibnamefont{and}
  \bibinfo{author}{\bibfnamefont{A.}~\bibnamefont{Wilson}},
  \bibinfo{journal}{Model. Simul. Mater. Sci.} \textbf{\bibinfo{volume}{25}},
  \bibinfo{pages}{015010} (\bibinfo{year}{2016}).

\bibitem[{\citenamefont{Dickel et~al.}(2018)\citenamefont{Dickel, Barrett,
  Carino, Baskes, and Horstemeyer}}]{Dickel2018}
\bibinfo{author}{\bibfnamefont{D.}~\bibnamefont{Dickel}},
  \bibinfo{author}{\bibfnamefont{C.~D.} \bibnamefont{Barrett}},
  \bibinfo{author}{\bibfnamefont{R.~L.} \bibnamefont{Carino}},
  \bibinfo{author}{\bibfnamefont{M.~I.} \bibnamefont{Baskes}},
  \bibnamefont{and} \bibinfo{author}{\bibfnamefont{M.~F.}
  \bibnamefont{Horstemeyer}}, \bibinfo{journal}{Model. Simul. Mater. Sci.}
  \textbf{\bibinfo{volume}{26}}, \bibinfo{pages}{065002}
  (\bibinfo{year}{2018}).

\bibitem[{\citenamefont{Mendelev et~al.}(2016)\citenamefont{Mendelev,
  Underwood, and Ackland}}]{Mendelev2016}
\bibinfo{author}{\bibfnamefont{M.~I.} \bibnamefont{Mendelev}},
  \bibinfo{author}{\bibfnamefont{T.~L.} \bibnamefont{Underwood}},
  \bibnamefont{and} \bibinfo{author}{\bibfnamefont{G.~J.}
  \bibnamefont{Ackland}}, \bibinfo{journal}{Phys. Rev. B}
  \textbf{\bibinfo{volume}{145}}, \bibinfo{pages}{154102}
  (\bibinfo{year}{2016}).

\bibitem[{\citenamefont{Kartamyshev et~al.}(2019)\citenamefont{Kartamyshev,
  Lipnitskii, Saveliev, Maksimenko, Nelasov, and Poletaev}}]{Kartamyshev2019}
\bibinfo{author}{\bibfnamefont{A.~I.} \bibnamefont{Kartamyshev}},
  \bibinfo{author}{\bibfnamefont{A.~G.} \bibnamefont{Lipnitskii}},
  \bibinfo{author}{\bibfnamefont{V.~N.} \bibnamefont{Saveliev}},
  \bibinfo{author}{\bibfnamefont{V.~N.} \bibnamefont{Maksimenko}},
  \bibinfo{author}{\bibfnamefont{I.~V.} \bibnamefont{Nelasov}},
  \bibnamefont{and} \bibinfo{author}{\bibfnamefont{D.~O.}
  \bibnamefont{Poletaev}}, \bibinfo{journal}{Comp. Mat. Sci.}
  \textbf{\bibinfo{volume}{160}}, \bibinfo{pages}{30} (\bibinfo{year}{2019}).

\bibitem[{\citenamefont{Hennig et~al.}(2008)\citenamefont{Hennig, Lenosky,
  Trinkle, Rudin, and Wilkins}}]{Hennig2008}
\bibinfo{author}{\bibfnamefont{R.~G.} \bibnamefont{Hennig}},
  \bibinfo{author}{\bibfnamefont{T.~J.} \bibnamefont{Lenosky}},
  \bibinfo{author}{\bibfnamefont{D.~R.} \bibnamefont{Trinkle}},
  \bibinfo{author}{\bibfnamefont{S.~P.} \bibnamefont{Rudin}}, \bibnamefont{and}
  \bibinfo{author}{\bibfnamefont{J.~W.} \bibnamefont{Wilkins}},
  \bibinfo{journal}{Phys. Rev. B} \textbf{\bibinfo{volume}{78}},
  \bibinfo{pages}{054121} (\bibinfo{year}{2008}).

\bibitem[{\citenamefont{Ehemann and Wilkins}(2017)}]{Ehemann2017}
\bibinfo{author}{\bibfnamefont{R.~C.} \bibnamefont{Ehemann}} \bibnamefont{and}
  \bibinfo{author}{\bibfnamefont{J.~W.} \bibnamefont{Wilkins}},
  \bibinfo{journal}{Phys. Rev. B} \textbf{\bibinfo{volume}{96}},
  \bibinfo{pages}{184105} (\bibinfo{year}{2017}).

\bibitem[{\citenamefont{Takahashi et~al.}(2017)\citenamefont{Takahashi, Seko,
  and Tanaka}}]{Takahashi2017}
\bibinfo{author}{\bibfnamefont{A.}~\bibnamefont{Takahashi}},
  \bibinfo{author}{\bibfnamefont{A.}~\bibnamefont{Seko}}, \bibnamefont{and}
  \bibinfo{author}{\bibfnamefont{I.}~\bibnamefont{Tanaka}},
  \bibinfo{journal}{Phys. Rev. Materials} \textbf{\bibinfo{volume}{1}},
  \bibinfo{pages}{063801} (\bibinfo{year}{2017}).

\bibitem[{\citenamefont{Rawat and Mitra}(2017)}]{Rawat2017}
\bibinfo{author}{\bibfnamefont{S.}~\bibnamefont{Rawat}} \bibnamefont{and}
  \bibinfo{author}{\bibfnamefont{N.}~\bibnamefont{Mitra}},
  \bibinfo{journal}{Comput. Mater. Sci.} \textbf{\bibinfo{volume}{126}},
  \bibinfo{pages}{228} (\bibinfo{year}{2017}).

\bibitem[{\citenamefont{Slater and Koster}(1954)}]{Slater1954}
\bibinfo{author}{\bibfnamefont{J.~C.} \bibnamefont{Slater}} \bibnamefont{and}
  \bibinfo{author}{\bibfnamefont{G.~F.} \bibnamefont{Koster}},
  \bibinfo{journal}{Phys. Rev.} \textbf{\bibinfo{volume}{94}},
  \bibinfo{pages}{1498} (\bibinfo{year}{1954}).

\bibitem[{\citenamefont{Ashcroft and Mermin}(1976)}]{Ashcroft1976}
\bibinfo{author}{\bibfnamefont{N.~W.} \bibnamefont{Ashcroft}} \bibnamefont{and}
  \bibinfo{author}{\bibfnamefont{N.~D.} \bibnamefont{Mermin}},
  \emph{\bibinfo{title}{Solid State Physics}} (\bibinfo{publisher}{Holt,
  Rinehart and Winston}, \bibinfo{year}{1976}).

\bibitem[{\citenamefont{Mehl and Papaconstantopoulos}(1996)}]{Mehl1996}
\bibinfo{author}{\bibfnamefont{M.~J.} \bibnamefont{Mehl}} \bibnamefont{and}
  \bibinfo{author}{\bibfnamefont{D.~A.} \bibnamefont{Papaconstantopoulos}},
  \bibinfo{journal}{Phys. Rev. B} \textbf{\bibinfo{volume}{54}},
  \bibinfo{pages}{4519} (\bibinfo{year}{1996}).

\bibitem[{\citenamefont{Mehl and Papaconstantopoulos}(2002)}]{Mehl2002}
\bibinfo{author}{\bibfnamefont{M.~J.} \bibnamefont{Mehl}} \bibnamefont{and}
  \bibinfo{author}{\bibfnamefont{D.~A.} \bibnamefont{Papaconstantopoulos}},
  \bibinfo{journal}{EPL} \textbf{\bibinfo{volume}{60}}, \bibinfo{pages}{248}
  (\bibinfo{year}{2002}).

\bibitem[{\citenamefont{Rudin et~al.}(2004)\citenamefont{Rudin, Jones, and
  Albers}}]{Rudin2004}
\bibinfo{author}{\bibfnamefont{S.~P.} \bibnamefont{Rudin}},
  \bibinfo{author}{\bibfnamefont{M.~D.} \bibnamefont{Jones}}, \bibnamefont{and}
  \bibinfo{author}{\bibfnamefont{R.~C.} \bibnamefont{Albers}},
  \bibinfo{journal}{Phys. Rev. B} \textbf{\bibinfo{volume}{69}},
  \bibinfo{pages}{094117} (\bibinfo{year}{2004}).

\bibitem[{\citenamefont{Trinkle et~al.}(2006)\citenamefont{Trinkle, Jones,
  Hennig, Rudin, Albers, and Wilkins}}]{Trinkle2006}
\bibinfo{author}{\bibfnamefont{D.~R.} \bibnamefont{Trinkle}},
  \bibinfo{author}{\bibfnamefont{M.~D.} \bibnamefont{Jones}},
  \bibinfo{author}{\bibfnamefont{R.~G.} \bibnamefont{Hennig}},
  \bibinfo{author}{\bibfnamefont{S.~P.} \bibnamefont{Rudin}},
  \bibinfo{author}{\bibfnamefont{R.~C.} \bibnamefont{Albers}},
  \bibnamefont{and} \bibinfo{author}{\bibfnamefont{J.~W.}
  \bibnamefont{Wilkins}}, \bibinfo{journal}{Phys. Rev. B}
  \textbf{\bibinfo{volume}{73}}, \bibinfo{pages}{094123}
  (\bibinfo{year}{2006}).

\bibitem[{\citenamefont{Margine et~al.}(2011)\citenamefont{Margine, Kolmogorov,
  Reese, Mrovec, Els{\"a}sser, Meyer, Drautz, and Pettifor}}]{Margine2011}
\bibinfo{author}{\bibfnamefont{E.~R.} \bibnamefont{Margine}},
  \bibinfo{author}{\bibfnamefont{A.~N.} \bibnamefont{Kolmogorov}},
  \bibinfo{author}{\bibfnamefont{M.}~\bibnamefont{Reese}},
  \bibinfo{author}{\bibfnamefont{M.}~\bibnamefont{Mrovec}},
  \bibinfo{author}{\bibfnamefont{C.}~\bibnamefont{Els{\"a}sser}},
  \bibinfo{author}{\bibfnamefont{B.}~\bibnamefont{Meyer}},
  \bibinfo{author}{\bibfnamefont{R.}~\bibnamefont{Drautz}}, \bibnamefont{and}
  \bibinfo{author}{\bibfnamefont{D.~G.} \bibnamefont{Pettifor}},
  \bibinfo{journal}{Phys. Rev. B} \textbf{\bibinfo{volume}{84}},
  \bibinfo{pages}{155120} (\bibinfo{year}{2011}).

\bibitem[{\citenamefont{Cawkwell et~al.}(2015)\citenamefont{Cawkwell, Coe,
  Yadav, Liu, and Niklasson}}]{Cawkwell2015}
\bibinfo{author}{\bibfnamefont{M.~J.} \bibnamefont{Cawkwell}},
  \bibinfo{author}{\bibfnamefont{J.~D.} \bibnamefont{Coe}},
  \bibinfo{author}{\bibfnamefont{S.~K.} \bibnamefont{Yadav}},
  \bibinfo{author}{\bibfnamefont{X.-Y.} \bibnamefont{Liu}}, \bibnamefont{and}
  \bibinfo{author}{\bibfnamefont{A.~M.~N.} \bibnamefont{Niklasson}},
  \bibinfo{journal}{J. Chem. Theory Comput.} \textbf{\bibinfo{volume}{11}},
  \bibinfo{pages}{2697} (\bibinfo{year}{2015}).

\bibitem[{\citenamefont{Pettifor}(1989)}]{Pettifor1989}
\bibinfo{author}{\bibfnamefont{D.~G.} \bibnamefont{Pettifor}},
  \bibinfo{journal}{Phys. Rev. Lett.} \textbf{\bibinfo{volume}{63}},
  \bibinfo{pages}{2480} (\bibinfo{year}{1989}).

\bibitem[{\citenamefont{Horsfield et~al.}(1996)\citenamefont{Horsfield,
  Bratkovsky, Fearn, Pettifor, and Aoki}}]{Horsfield1996}
\bibinfo{author}{\bibfnamefont{A.~P.} \bibnamefont{Horsfield}},
  \bibinfo{author}{\bibfnamefont{A.~M.} \bibnamefont{Bratkovsky}},
  \bibinfo{author}{\bibfnamefont{M.}~\bibnamefont{Fearn}},
  \bibinfo{author}{\bibfnamefont{D.~G.} \bibnamefont{Pettifor}},
  \bibnamefont{and} \bibinfo{author}{\bibfnamefont{M.}~\bibnamefont{Aoki}},
  \bibinfo{journal}{Phys. Rev. B} \textbf{\bibinfo{volume}{53}},
  \bibinfo{pages}{12694} (\bibinfo{year}{1996}).

\bibitem[{\citenamefont{Pettifor et~al.}(2002)\citenamefont{Pettifor, Oleinik,
  Nguyen-Manh, and Vitek}}]{Pettifor2002}
\bibinfo{author}{\bibfnamefont{D.~G.} \bibnamefont{Pettifor}},
  \bibinfo{author}{\bibfnamefont{I.~I.} \bibnamefont{Oleinik}},
  \bibinfo{author}{\bibfnamefont{D.}~\bibnamefont{Nguyen-Manh}},
  \bibnamefont{and} \bibinfo{author}{\bibfnamefont{V.}~\bibnamefont{Vitek}},
  \bibinfo{journal}{Comp. Mat. Sci.} \textbf{\bibinfo{volume}{23}},
  \bibinfo{pages}{33} (\bibinfo{year}{2002}).

\bibitem[{\citenamefont{Drautz and Pettifor}(2006)}]{Drautz2006}
\bibinfo{author}{\bibfnamefont{R.}~\bibnamefont{Drautz}} \bibnamefont{and}
  \bibinfo{author}{\bibfnamefont{D.~G.} \bibnamefont{Pettifor}},
  \bibinfo{journal}{Phys. Rev. B} \textbf{\bibinfo{volume}{74}},
  \bibinfo{pages}{174117} (\bibinfo{year}{2006}).

\bibitem[{\citenamefont{Girshick
  et~al.}(1998{\natexlab{a}})\citenamefont{Girshick, Bratkovsky, Pettifor, and
  Vitek}}]{Girshick1998}
\bibinfo{author}{\bibfnamefont{A.}~\bibnamefont{Girshick}},
  \bibinfo{author}{\bibfnamefont{A.~M.} \bibnamefont{Bratkovsky}},
  \bibinfo{author}{\bibfnamefont{D.~G.} \bibnamefont{Pettifor}},
  \bibnamefont{and} \bibinfo{author}{\bibfnamefont{V.}~\bibnamefont{Vitek}},
  \bibinfo{journal}{Phylos. Mag. A} \textbf{\bibinfo{volume}{77}},
  \bibinfo{pages}{981} (\bibinfo{year}{1998}{\natexlab{a}}).

\bibitem[{\citenamefont{Girshick
  et~al.}(1998{\natexlab{b}})\citenamefont{Girshick, Pettifor, and
  Vitek}}]{Girshick1998a}
\bibinfo{author}{\bibfnamefont{A.}~\bibnamefont{Girshick}},
  \bibinfo{author}{\bibfnamefont{D.~G.} \bibnamefont{Pettifor}},
  \bibnamefont{and} \bibinfo{author}{\bibfnamefont{V.}~\bibnamefont{Vitek}},
  \bibinfo{journal}{Phylos. Mag. A} \textbf{\bibinfo{volume}{77}},
  \bibinfo{pages}{999} (\bibinfo{year}{1998}{\natexlab{b}}).

\bibitem[{\citenamefont{Nguyen-Manh et~al.}(2000)\citenamefont{Nguyen-Manh,
  Pettifor, and Vitek}}]{Nguyen-Manh2000}
\bibinfo{author}{\bibfnamefont{D.}~\bibnamefont{Nguyen-Manh}},
  \bibinfo{author}{\bibfnamefont{D.~G.} \bibnamefont{Pettifor}},
  \bibnamefont{and} \bibinfo{author}{\bibfnamefont{V.}~\bibnamefont{Vitek}},
  \bibinfo{journal}{Phys. Rev. Lett.} \textbf{\bibinfo{volume}{85}},
  \bibinfo{pages}{4136} (\bibinfo{year}{2000}).

\bibitem[{\citenamefont{L{\"o}wdin}(1950)}]{Loewdin1950}
\bibinfo{author}{\bibfnamefont{P.-O.} \bibnamefont{L{\"o}wdin}},
  \bibinfo{journal}{J. Chem. Phys.} \textbf{\bibinfo{volume}{18}},
  \bibinfo{pages}{365} (\bibinfo{year}{1950}).

\bibitem[{\citenamefont{Urban et~al.}(2011)\citenamefont{Urban, Reese, Mrovec,
  Els{\"a}sser, and Meyer}}]{Urban2011}
\bibinfo{author}{\bibfnamefont{A.}~\bibnamefont{Urban}},
  \bibinfo{author}{\bibfnamefont{M.}~\bibnamefont{Reese}},
  \bibinfo{author}{\bibfnamefont{M.}~\bibnamefont{Mrovec}},
  \bibinfo{author}{\bibfnamefont{C.}~\bibnamefont{Els{\"a}sser}},
  \bibnamefont{and} \bibinfo{author}{\bibfnamefont{B.}~\bibnamefont{Meyer}},
  \bibinfo{journal}{Phys. Rev. B} \textbf{\bibinfo{volume}{84}},
  \bibinfo{pages}{155119} (\bibinfo{year}{2011}).

\bibitem[{\citenamefont{Pettifor}(1970)}]{Pettifor1970}
\bibinfo{author}{\bibfnamefont{D.~G.} \bibnamefont{Pettifor}},
  \bibinfo{journal}{J. Phys. C: Solid State Phys.}
  \textbf{\bibinfo{volume}{3}}, \bibinfo{pages}{367} (\bibinfo{year}{1970}).

\bibitem[{\citenamefont{Pettifor}(1986)}]{Pettifor1986}
\bibinfo{author}{\bibfnamefont{D.~G.} \bibnamefont{Pettifor}},
  \bibinfo{journal}{J. Phys. C: Solid State Phys.}
  \textbf{\bibinfo{volume}{19}}, \bibinfo{pages}{285} (\bibinfo{year}{1986}).

\bibitem[{\citenamefont{Seiser et~al.}(2011)\citenamefont{Seiser,
  Hammerschmidt, Kolmogorov, Drautz, and Pettifor}}]{Seiser2011}
\bibinfo{author}{\bibfnamefont{B.}~\bibnamefont{Seiser}},
  \bibinfo{author}{\bibfnamefont{T.}~\bibnamefont{Hammerschmidt}},
  \bibinfo{author}{\bibfnamefont{A.~N.} \bibnamefont{Kolmogorov}},
  \bibinfo{author}{\bibfnamefont{R.}~\bibnamefont{Drautz}}, \bibnamefont{and}
  \bibinfo{author}{\bibfnamefont{D.~G.} \bibnamefont{Pettifor}},
  \bibinfo{journal}{Phys. Rev. B} \textbf{\bibinfo{volume}{83}},
  \bibinfo{pages}{224116} (\bibinfo{year}{2011}).

\bibitem[{\citenamefont{Pettifor}(1977)}]{Pettifor1977}
\bibinfo{author}{\bibfnamefont{D.~G.} \bibnamefont{Pettifor}},
  \bibinfo{journal}{J. Phys. F: Met. Phys.} \textbf{\bibinfo{volume}{7}},
  \bibinfo{pages}{613} (\bibinfo{year}{1977}).

\bibitem[{\citenamefont{Andersen et~al.}(1978)\citenamefont{Andersen, Klose,
  and Nohl}}]{Andersen1978}
\bibinfo{author}{\bibfnamefont{O.~K.} \bibnamefont{Andersen}},
  \bibinfo{author}{\bibfnamefont{W.}~\bibnamefont{Klose}}, \bibnamefont{and}
  \bibinfo{author}{\bibfnamefont{H.}~\bibnamefont{Nohl}},
  \bibinfo{journal}{Phys. Rev. B} \textbf{\bibinfo{volume}{17}},
  \bibinfo{pages}{1209} (\bibinfo{year}{1978}).

\bibitem[{\citenamefont{Pettifor}(1995)}]{Pettifor1995}
\bibinfo{author}{\bibfnamefont{D.~G.} \bibnamefont{Pettifor}},
  \emph{\bibinfo{title}{Bonding and structure of molecules and solids}}
  (\bibinfo{publisher}{Oxford University Press}, \bibinfo{year}{1995}).

\bibitem[{\citenamefont{Mrovec et~al.}(2004)\citenamefont{Mrovec, Nguyen-Manh,
  Pettifor, and Vitek}}]{Mrovec2004}
\bibinfo{author}{\bibfnamefont{M.}~\bibnamefont{Mrovec}},
  \bibinfo{author}{\bibfnamefont{D.}~\bibnamefont{Nguyen-Manh}},
  \bibinfo{author}{\bibfnamefont{D.~G.} \bibnamefont{Pettifor}},
  \bibnamefont{and} \bibinfo{author}{\bibfnamefont{V.}~\bibnamefont{Vitek}},
  \bibinfo{journal}{Phys. Rev. B} \textbf{\bibinfo{volume}{69}},
  \bibinfo{pages}{094115} (\bibinfo{year}{2004}).

\bibitem[{\citenamefont{{\v{C}}{\'a}k et~al.}(2014)\citenamefont{{\v{C}}{\'a}k,
  Hammerschmidt, Rogal, Vitek, and Drautz}}]{Cak2014}
\bibinfo{author}{\bibfnamefont{M.}~\bibnamefont{{\v{C}}{\'a}k}},
  \bibinfo{author}{\bibfnamefont{T.}~\bibnamefont{Hammerschmidt}},
  \bibinfo{author}{\bibfnamefont{J.}~\bibnamefont{Rogal}},
  \bibinfo{author}{\bibfnamefont{V.}~\bibnamefont{Vitek}}, \bibnamefont{and}
  \bibinfo{author}{\bibfnamefont{R.}~\bibnamefont{Drautz}},
  \bibinfo{journal}{J. Phys.: Condens. Matter} \textbf{\bibinfo{volume}{26}},
  \bibinfo{pages}{195501} (\bibinfo{year}{2014}).

\bibitem[{\citenamefont{Lin et~al.}(2014)\citenamefont{Lin, Mrovec, and
  Vitek}}]{Lin2014}
\bibinfo{author}{\bibfnamefont{Y.~S.} \bibnamefont{Lin}},
  \bibinfo{author}{\bibfnamefont{M.}~\bibnamefont{Mrovec}}, \bibnamefont{and}
  \bibinfo{author}{\bibfnamefont{V.}~\bibnamefont{Vitek}},
  \bibinfo{journal}{Model. Simul. Mater. Sci.} \textbf{\bibinfo{volume}{22}},
  \bibinfo{pages}{034002} (\bibinfo{year}{2014}).

\bibitem[{\citenamefont{Cawkwell et~al.}(2005)\citenamefont{Cawkwell,
  Nguyen-Manh, Woodward, Pettifor, and Vitek}}]{Cawkwell2005}
\bibinfo{author}{\bibfnamefont{M.~J.} \bibnamefont{Cawkwell}},
  \bibinfo{author}{\bibfnamefont{D.}~\bibnamefont{Nguyen-Manh}},
  \bibinfo{author}{\bibfnamefont{C.}~\bibnamefont{Woodward}},
  \bibinfo{author}{\bibfnamefont{D.~G.} \bibnamefont{Pettifor}},
  \bibnamefont{and} \bibinfo{author}{\bibfnamefont{V.}~\bibnamefont{Vitek}},
  \bibinfo{journal}{Science} \textbf{\bibinfo{volume}{309}},
  \bibinfo{pages}{1059} (\bibinfo{year}{2005}).

\bibitem[{\citenamefont{Cawkwell et~al.}(2006)\citenamefont{Cawkwell,
  Nguyen-Manh, Pettifor, and Vitek}}]{Cawkwell2006}
\bibinfo{author}{\bibfnamefont{M.~J.} \bibnamefont{Cawkwell}},
  \bibinfo{author}{\bibfnamefont{D.}~\bibnamefont{Nguyen-Manh}},
  \bibinfo{author}{\bibfnamefont{D.~G.} \bibnamefont{Pettifor}},
  \bibnamefont{and} \bibinfo{author}{\bibfnamefont{V.}~\bibnamefont{Vitek}},
  \bibinfo{journal}{Phys. Rev. B} \textbf{\bibinfo{volume}{73}},
  \bibinfo{pages}{064104} (\bibinfo{year}{2006}).

\bibitem[{\citenamefont{Mrovec et~al.}(2007)\citenamefont{Mrovec, Gr{\"o}ger,
  Bailey, Nguyen-Manh, Els{\"a}sser, and Vitek}}]{Mrovec2007}
\bibinfo{author}{\bibfnamefont{M.}~\bibnamefont{Mrovec}},
  \bibinfo{author}{\bibfnamefont{R.}~\bibnamefont{Gr{\"o}ger}},
  \bibinfo{author}{\bibfnamefont{A.~G.} \bibnamefont{Bailey}},
  \bibinfo{author}{\bibfnamefont{D.}~\bibnamefont{Nguyen-Manh}},
  \bibinfo{author}{\bibfnamefont{C.}~\bibnamefont{Els{\"a}sser}},
  \bibnamefont{and} \bibinfo{author}{\bibfnamefont{V.}~\bibnamefont{Vitek}},
  \bibinfo{journal}{Phys. Rev. B} \textbf{\bibinfo{volume}{75}},
  \bibinfo{pages}{104119} (\bibinfo{year}{2007}).

\bibitem[{\citenamefont{Mrovec et~al.}(2011)\citenamefont{Mrovec, Nguyen-Manh,
  Els{\"a}sser, and Gumbsch}}]{Mrovec2011}
\bibinfo{author}{\bibfnamefont{M.}~\bibnamefont{Mrovec}},
  \bibinfo{author}{\bibfnamefont{D.}~\bibnamefont{Nguyen-Manh}},
  \bibinfo{author}{\bibfnamefont{C.}~\bibnamefont{Els{\"a}sser}},
  \bibnamefont{and} \bibinfo{author}{\bibfnamefont{P.}~\bibnamefont{Gumbsch}},
  \bibinfo{journal}{Phys. Rev. Lett.} \textbf{\bibinfo{volume}{106}},
  \bibinfo{pages}{246402} (\bibinfo{year}{2011}).

\bibitem[{\citenamefont{Lin et~al.}(2016)\citenamefont{Lin, Mrovec, and
  Vitek}}]{Lin2016}
\bibinfo{author}{\bibfnamefont{Y.~S.} \bibnamefont{Lin}},
  \bibinfo{author}{\bibfnamefont{M.}~\bibnamefont{Mrovec}}, \bibnamefont{and}
  \bibinfo{author}{\bibfnamefont{V.}~\bibnamefont{Vitek}},
  \bibinfo{journal}{Phys. Rev. B} \textbf{\bibinfo{volume}{93}},
  \bibinfo{pages}{214107} (\bibinfo{year}{2016}).

\bibitem[{\citenamefont{Drain et~al.}(2014)\citenamefont{Drain, Drautz, and
  Pettifor}}]{Drain2014}
\bibinfo{author}{\bibfnamefont{J.~F.} \bibnamefont{Drain}},
  \bibinfo{author}{\bibfnamefont{R.}~\bibnamefont{Drautz}}, \bibnamefont{and}
  \bibinfo{author}{\bibfnamefont{D.~G.} \bibnamefont{Pettifor}},
  \bibinfo{journal}{Phys. Rev. B} \textbf{\bibinfo{volume}{89}},
  \bibinfo{pages}{134102} (\bibinfo{year}{2014}).

\bibitem[{\citenamefont{Cyrot-Lackmann}(1967)}]{Cyrot-Lackmann1967}
\bibinfo{author}{\bibfnamefont{F.}~\bibnamefont{Cyrot-Lackmann}},
  \bibinfo{journal}{Adv. Phys.} \textbf{\bibinfo{volume}{16}},
  \bibinfo{pages}{393} (\bibinfo{year}{1967}).

\bibitem[{\citenamefont{Jenke et~al.}(2018)\citenamefont{Jenke, Subramanyam,
  Densow, Hammerschmidt, Pettifor, and Drautz}}]{Jenke2018}
\bibinfo{author}{\bibfnamefont{J.}~\bibnamefont{Jenke}},
  \bibinfo{author}{\bibfnamefont{A.}~\bibnamefont{Subramanyam}},
  \bibinfo{author}{\bibfnamefont{M.}~\bibnamefont{Densow}},
  \bibinfo{author}{\bibfnamefont{T.}~\bibnamefont{Hammerschmidt}},
  \bibinfo{author}{\bibfnamefont{D.}~\bibnamefont{Pettifor}}, \bibnamefont{and}
  \bibinfo{author}{\bibfnamefont{R.}~\bibnamefont{Drautz}},
  \bibinfo{journal}{Phys. Rev. B} \textbf{\bibinfo{volume}{98}},
  \bibinfo{pages}{144102} (\bibinfo{year}{2018}).

\bibitem[{\citenamefont{Seiser et~al.}(2013)\citenamefont{Seiser, Pettifor, and
  Drautz}}]{Seiser2013}
\bibinfo{author}{\bibfnamefont{B.}~\bibnamefont{Seiser}},
  \bibinfo{author}{\bibfnamefont{D.~G.} \bibnamefont{Pettifor}},
  \bibnamefont{and} \bibinfo{author}{\bibfnamefont{R.}~\bibnamefont{Drautz}},
  \bibinfo{journal}{Phys. Rev. B} \textbf{\bibinfo{volume}{87}},
  \bibinfo{pages}{094105} (\bibinfo{year}{2013}).

\bibitem[{\citenamefont{Drautz and Pettifor}(2011)}]{Drautz2011}
\bibinfo{author}{\bibfnamefont{R.}~\bibnamefont{Drautz}} \bibnamefont{and}
  \bibinfo{author}{\bibfnamefont{D.~G.} \bibnamefont{Pettifor}},
  \bibinfo{journal}{Phys. Rev. B} \textbf{\bibinfo{volume}{84}},
  \bibinfo{pages}{214114} (\bibinfo{year}{2011}).

\bibitem[{\citenamefont{Hammerschmidt et~al.}(2019)\citenamefont{Hammerschmidt,
  Seiser, Ford, Ladines, Schreiber, Wang, Jenke, Lysogorskiy, Teijeiro, Mrovec
  et~al.}}]{Hammerschmidt2018}
\bibinfo{author}{\bibfnamefont{T.}~\bibnamefont{Hammerschmidt}},
  \bibinfo{author}{\bibfnamefont{B.}~\bibnamefont{Seiser}},
  \bibinfo{author}{\bibfnamefont{M.~E.} \bibnamefont{Ford}},
  \bibinfo{author}{\bibfnamefont{A.~N.} \bibnamefont{Ladines}},
  \bibinfo{author}{\bibfnamefont{S.}~\bibnamefont{Schreiber}},
  \bibinfo{author}{\bibfnamefont{N.}~\bibnamefont{Wang}},
  \bibinfo{author}{\bibfnamefont{J.}~\bibnamefont{Jenke}},
  \bibinfo{author}{\bibfnamefont{Y.}~\bibnamefont{Lysogorskiy}},
  \bibinfo{author}{\bibfnamefont{C.}~\bibnamefont{Teijeiro}},
  \bibinfo{author}{\bibfnamefont{M.}~\bibnamefont{Mrovec}},
  \bibnamefont{et~al.}, \bibinfo{journal}{Comput. Phys. Commun.}
  \textbf{\bibinfo{volume}{235}}, \bibinfo{pages}{221} (\bibinfo{year}{2019}).

\bibitem[{\citenamefont{Madsen et~al.}(2011)\citenamefont{Madsen, McEniry, and
  Drautz}}]{Madsen2011}
\bibinfo{author}{\bibfnamefont{G.~K.~H.} \bibnamefont{Madsen}},
  \bibinfo{author}{\bibfnamefont{E.~J.} \bibnamefont{McEniry}},
  \bibnamefont{and} \bibinfo{author}{\bibfnamefont{R.}~\bibnamefont{Drautz}},
  \bibinfo{journal}{Phys. Rev. B} \textbf{\bibinfo{volume}{83}},
  \bibinfo{pages}{184119} (\bibinfo{year}{2011}).

\bibitem[{\citenamefont{Finnis and Sinclair}(1984)}]{Finnis1984}
\bibinfo{author}{\bibfnamefont{M.~W.} \bibnamefont{Finnis}} \bibnamefont{and}
  \bibinfo{author}{\bibfnamefont{J.~E.} \bibnamefont{Sinclair}},
  \bibinfo{journal}{Phylos. Mag. A} \textbf{\bibinfo{volume}{50}},
  \bibinfo{pages}{45} (\bibinfo{year}{1984}).

\bibitem[{\citenamefont{Lysogorskiy et~al.}(2019)\citenamefont{Lysogorskiy,
  Hammerschmidt, Janssen, Neugebauer, and Drautz}}]{Lysogorskiy2019}
\bibinfo{author}{\bibfnamefont{Y.~V.} \bibnamefont{Lysogorskiy}},
  \bibinfo{author}{\bibfnamefont{T.}~\bibnamefont{Hammerschmidt}},
  \bibinfo{author}{\bibfnamefont{J.}~\bibnamefont{Janssen}},
  \bibinfo{author}{\bibfnamefont{J.}~\bibnamefont{Neugebauer}},
  \bibnamefont{and} \bibinfo{author}{\bibfnamefont{R.}~\bibnamefont{Drautz}},
  \bibinfo{journal}{Model. Simul. Mater. Sci.} \textbf{\bibinfo{volume}{27}},
  \bibinfo{pages}{025007} (\bibinfo{year}{2019}).

\bibitem[{\citenamefont{Kresse and Hafner}(1993)}]{Kresse1993}
\bibinfo{author}{\bibfnamefont{G.}~\bibnamefont{Kresse}} \bibnamefont{and}
  \bibinfo{author}{\bibfnamefont{J.}~\bibnamefont{Hafner}},
  \bibinfo{journal}{Phys. Rev. B} \textbf{\bibinfo{volume}{47}},
  \bibinfo{pages}{558} (\bibinfo{year}{1993}).

\bibitem[{\citenamefont{Kresse and
  Furthm{\"u}ller}(1996{\natexlab{a}})}]{Kresse1996}
\bibinfo{author}{\bibfnamefont{G.}~\bibnamefont{Kresse}} \bibnamefont{and}
  \bibinfo{author}{\bibfnamefont{J.}~\bibnamefont{Furthm{\"u}ller}},
  \bibinfo{journal}{Comput. Mat. Sci.} \textbf{\bibinfo{volume}{6}},
  \bibinfo{pages}{15} (\bibinfo{year}{1996}{\natexlab{a}}).

\bibitem[{\citenamefont{Kresse and
  Furthm{\"u}ller}(1996{\natexlab{b}})}]{Kresse1996a}
\bibinfo{author}{\bibfnamefont{G.}~\bibnamefont{Kresse}} \bibnamefont{and}
  \bibinfo{author}{\bibfnamefont{J.}~\bibnamefont{Furthm{\"u}ller}},
  \bibinfo{journal}{Phys. Rev. B} \textbf{\bibinfo{volume}{54}},
  \bibinfo{pages}{11169} (\bibinfo{year}{1996}{\natexlab{b}}).

\bibitem[{\citenamefont{Janssen et~al.}(2019)\citenamefont{Janssen,
  Surendralal, Lysogorskiy, Todorova, Hickel, Drautz, and
  Neugebauer}}]{Janssen2018}
\bibinfo{author}{\bibfnamefont{J.}~\bibnamefont{Janssen}},
  \bibinfo{author}{\bibfnamefont{S.}~\bibnamefont{Surendralal}},
  \bibinfo{author}{\bibfnamefont{Y.}~\bibnamefont{Lysogorskiy}},
  \bibinfo{author}{\bibfnamefont{M.}~\bibnamefont{Todorova}},
  \bibinfo{author}{\bibfnamefont{T.}~\bibnamefont{Hickel}},
  \bibinfo{author}{\bibfnamefont{R.}~\bibnamefont{Drautz}}, \bibnamefont{and}
  \bibinfo{author}{\bibfnamefont{J.}~\bibnamefont{Neugebauer}},
  \bibinfo{journal}{Comp. Mat. Sci.} \textbf{\bibinfo{volume}{163}},
  \bibinfo{pages}{24} (\bibinfo{year}{2019}).

\bibitem[{\citenamefont{Bl{\"o}chl}(1994)}]{Bloechl1994}
\bibinfo{author}{\bibfnamefont{P.~E.} \bibnamefont{Bl{\"o}chl}},
  \bibinfo{journal}{Phys. Rev. B} \textbf{\bibinfo{volume}{50}},
  \bibinfo{pages}{17953} (\bibinfo{year}{1994}).

\bibitem[{\citenamefont{Kresse and Joubert}(1999)}]{Kresse1999}
\bibinfo{author}{\bibfnamefont{G.}~\bibnamefont{Kresse}} \bibnamefont{and}
  \bibinfo{author}{\bibfnamefont{D.}~\bibnamefont{Joubert}},
  \bibinfo{journal}{Phys. Rev. B} \textbf{\bibinfo{volume}{59}},
  \bibinfo{pages}{1758} (\bibinfo{year}{1999}).

\bibitem[{\citenamefont{Perdew et~al.}(1996)\citenamefont{Perdew, Burke, and
  Ernzerhof}}]{Perdew1996}
\bibinfo{author}{\bibfnamefont{J.~P.} \bibnamefont{Perdew}},
  \bibinfo{author}{\bibfnamefont{K.}~\bibnamefont{Burke}}, \bibnamefont{and}
  \bibinfo{author}{\bibfnamefont{M.}~\bibnamefont{Ernzerhof}},
  \bibinfo{journal}{Phys. Rev. Lett.} \textbf{\bibinfo{volume}{77}},
  \bibinfo{pages}{3865} (\bibinfo{year}{1996}).

\bibitem[{\citenamefont{Baldereschi}(1973)}]{Baldereschi1973}
\bibinfo{author}{\bibfnamefont{A.}~\bibnamefont{Baldereschi}},
  \bibinfo{journal}{Phys. Rev. B} \textbf{\bibinfo{volume}{7}},
  \bibinfo{pages}{5212} (\bibinfo{year}{1973}).

\bibitem[{\citenamefont{Monkhorst and Pack}(1976)}]{Monkhorst1976}
\bibinfo{author}{\bibfnamefont{H.~J.} \bibnamefont{Monkhorst}}
  \bibnamefont{and} \bibinfo{author}{\bibfnamefont{J.~D.} \bibnamefont{Pack}},
  \bibinfo{journal}{Phys. Rev. B} \textbf{\bibinfo{volume}{13}},
  \bibinfo{pages}{5188} (\bibinfo{year}{1976}).

\bibitem[{\citenamefont{Methfessel and Paxton}(1989)}]{Methfessel1989}
\bibinfo{author}{\bibfnamefont{M.~P. A.~T.} \bibnamefont{Methfessel}}
  \bibnamefont{and} \bibinfo{author}{\bibfnamefont{A.~T.}
  \bibnamefont{Paxton}}, \bibinfo{journal}{Phys. Rev. B}
  \textbf{\bibinfo{volume}{40}}, \bibinfo{pages}{3616} (\bibinfo{year}{1989}).

\bibitem[{\citenamefont{Dacosta et~al.}(1986)\citenamefont{Dacosta, Nielsen,
  and Kunc}}]{DaCosta1986}
\bibinfo{author}{\bibfnamefont{P.~G.} \bibnamefont{Dacosta}},
  \bibinfo{author}{\bibfnamefont{O.~H.} \bibnamefont{Nielsen}},
  \bibnamefont{and} \bibinfo{author}{\bibfnamefont{K.}~\bibnamefont{Kunc}},
  \bibinfo{journal}{J. Phys. C:} \textbf{\bibinfo{volume}{19}},
  \bibinfo{pages}{3163} (\bibinfo{year}{1986}).

\bibitem[{\citenamefont{Golesorkhtabar
  et~al.}(2013)\citenamefont{Golesorkhtabar, Pavone, Spitaler, Puschnig, and
  Draxl}}]{Golesorkhtabar2013}
\bibinfo{author}{\bibfnamefont{R.}~\bibnamefont{Golesorkhtabar}},
  \bibinfo{author}{\bibfnamefont{P.}~\bibnamefont{Pavone}},
  \bibinfo{author}{\bibfnamefont{J.}~\bibnamefont{Spitaler}},
  \bibinfo{author}{\bibfnamefont{P.}~\bibnamefont{Puschnig}}, \bibnamefont{and}
  \bibinfo{author}{\bibfnamefont{C.}~\bibnamefont{Draxl}},
  \bibinfo{journal}{Comput. Phys. Commun.} \textbf{\bibinfo{volume}{184}},
  \bibinfo{pages}{1861} (\bibinfo{year}{2013}).

\bibitem[{\citenamefont{Togo and Tanaka}(2015)}]{Togo2015}
\bibinfo{author}{\bibfnamefont{A.}~\bibnamefont{Togo}} \bibnamefont{and}
  \bibinfo{author}{\bibfnamefont{I.}~\bibnamefont{Tanaka}},
  \bibinfo{journal}{Scr. Mater.} \textbf{\bibinfo{volume}{108}},
  \bibinfo{pages}{1} (\bibinfo{year}{2015}).

\bibitem[{\citenamefont{Levenberg}(1944)}]{Levenberg1944}
\bibinfo{author}{\bibfnamefont{K.}~\bibnamefont{Levenberg}},
  \bibinfo{journal}{Q. Appl. Math.} \textbf{\bibinfo{volume}{2}},
  \bibinfo{pages}{164} (\bibinfo{year}{1944}).

\bibitem[{\citenamefont{Marquardt}(1963)}]{Marquardt1963}
\bibinfo{author}{\bibfnamefont{D.~W.} \bibnamefont{Marquardt}},
  \bibinfo{journal}{J. Soc. Ind. Appl. Math.} \textbf{\bibinfo{volume}{11}},
  \bibinfo{pages}{431} (\bibinfo{year}{1963}).

\bibitem[{\citenamefont{Ladines}(2019)}]{Ladines}
\bibinfo{author}{\bibfnamefont{A.}~\bibnamefont{Ladines}}
  (\bibinfo{year}{2019}), \bibinfo{note}{manuscript in preparation}.

\bibitem[{\citenamefont{Jenke}(2019)}]{Jenke2019}
\bibinfo{author}{\bibfnamefont{J.}~\bibnamefont{Jenke}}, Ph.D. thesis,
  \bibinfo{school}{Ruhr-Universit{\"a}t Bochum} (\bibinfo{year}{2019}).

\bibitem[{Sup()}]{SuppMat}
\emph{\bibinfo{title}{{See Supplemental Material}}}.

\bibitem[{\citenamefont{Mrovec}(2019)}]{Mrovec:unpublished}
\bibinfo{author}{\bibfnamefont{M.}~\bibnamefont{Mrovec}}
  (\bibinfo{year}{2019}), \bibinfo{note}{unpublished}.

\bibitem[{\citenamefont{Stassis et~al.}(1979)\citenamefont{Stassis, Arch,
  Harmon, and Wakabayashi}}]{Stassis1979}
\bibinfo{author}{\bibfnamefont{C.}~\bibnamefont{Stassis}},
  \bibinfo{author}{\bibfnamefont{D.}~\bibnamefont{Arch}},
  \bibinfo{author}{\bibfnamefont{B.~N.} \bibnamefont{Harmon}},
  \bibnamefont{and}
  \bibinfo{author}{\bibfnamefont{N.}~\bibnamefont{Wakabayashi}},
  \bibinfo{journal}{Phys. Rev. B} \textbf{\bibinfo{volume}{19}},
  \bibinfo{pages}{181} (\bibinfo{year}{1979}).

\bibitem[{\citenamefont{Petry et~al.}(1991)\citenamefont{Petry, Heiming,
  Trampenau, Alba, Herzig, Schober, and Vogl}}]{Petry1991}
\bibinfo{author}{\bibfnamefont{W.}~\bibnamefont{Petry}},
  \bibinfo{author}{\bibfnamefont{A.}~\bibnamefont{Heiming}},
  \bibinfo{author}{\bibfnamefont{J.}~\bibnamefont{Trampenau}},
  \bibinfo{author}{\bibfnamefont{M.}~\bibnamefont{Alba}},
  \bibinfo{author}{\bibfnamefont{C.}~\bibnamefont{Herzig}},
  \bibinfo{author}{\bibfnamefont{H.~R.} \bibnamefont{Schober}},
  \bibnamefont{and} \bibinfo{author}{\bibfnamefont{G.}~\bibnamefont{Vogl}},
  \bibinfo{journal}{Phys. Rev. B} \textbf{\bibinfo{volume}{43}},
  \bibinfo{pages}{10933} (\bibinfo{year}{1991}).

\bibitem[{\citenamefont{Tane et~al.}(2013)\citenamefont{Tane, Okuda, Todaka,
  Ogi, and Nagakubo}}]{Tane2013}
\bibinfo{author}{\bibfnamefont{M.}~\bibnamefont{Tane}},
  \bibinfo{author}{\bibfnamefont{Y.}~\bibnamefont{Okuda}},
  \bibinfo{author}{\bibfnamefont{Y.}~\bibnamefont{Todaka}},
  \bibinfo{author}{\bibfnamefont{H.}~\bibnamefont{Ogi}}, \bibnamefont{and}
  \bibinfo{author}{\bibfnamefont{A.}~\bibnamefont{Nagakubo}},
  \bibinfo{journal}{Acta Mat.} \textbf{\bibinfo{volume}{61}},
  \bibinfo{pages}{7543} (\bibinfo{year}{2013}).

\bibitem[{\citenamefont{Fisher and Renken}(1964)}]{Fisher1964}
\bibinfo{author}{\bibfnamefont{E.~S.} \bibnamefont{Fisher}} \bibnamefont{and}
  \bibinfo{author}{\bibfnamefont{C.~J.} \bibnamefont{Renken}},
  \bibinfo{journal}{Phys. Rev.} \textbf{\bibinfo{volume}{64}},
  \bibinfo{pages}{A482} (\bibinfo{year}{1964}).

\bibitem[{\citenamefont{Ledbetter et~al.}(2004)\citenamefont{Ledbetter, Ogi,
  Kai, Kim, and Hirao}}]{Ledbetter2004}
\bibinfo{author}{\bibfnamefont{H.}~\bibnamefont{Ledbetter}},
  \bibinfo{author}{\bibfnamefont{H.}~\bibnamefont{Ogi}},
  \bibinfo{author}{\bibfnamefont{S.}~\bibnamefont{Kai}},
  \bibinfo{author}{\bibfnamefont{S.}~\bibnamefont{Kim}}, \bibnamefont{and}
  \bibinfo{author}{\bibfnamefont{M.}~\bibnamefont{Hirao}}, \bibinfo{journal}{J.
  Appl. Phys.} \textbf{\bibinfo{volume}{95}}, \bibinfo{pages}{4642}
  (\bibinfo{year}{2004}).

\bibitem[{\citenamefont{Paidar et~al.}(1999)\citenamefont{Paidar, Wang, Sob,
  and Vitek}}]{Paidar1999}
\bibinfo{author}{\bibfnamefont{V.}~\bibnamefont{Paidar}},
  \bibinfo{author}{\bibfnamefont{L.~G.} \bibnamefont{Wang}},
  \bibinfo{author}{\bibfnamefont{M.}~\bibnamefont{Sob}}, \bibnamefont{and}
  \bibinfo{author}{\bibfnamefont{V.}~\bibnamefont{Vitek}},
  \bibinfo{journal}{Modelling Simul. Mater. Sci. Eng.}
  \textbf{\bibinfo{volume}{7}}, \bibinfo{pages}{369} (\bibinfo{year}{1999}).

\bibitem[{\citenamefont{Luo et~al.}(2002)\citenamefont{Luo, Roundy, Cohen, and
  Morris~Jr}}]{Luo2002}
\bibinfo{author}{\bibfnamefont{W.}~\bibnamefont{Luo}},
  \bibinfo{author}{\bibfnamefont{D.}~\bibnamefont{Roundy}},
  \bibinfo{author}{\bibfnamefont{M.~L.} \bibnamefont{Cohen}}, \bibnamefont{and}
  \bibinfo{author}{\bibfnamefont{J.~W.} \bibnamefont{Morris~Jr}},
  \bibinfo{journal}{Phys. Rev. B} \textbf{\bibinfo{volume}{66}},
  \bibinfo{pages}{094110} (\bibinfo{year}{2002}).

\bibitem[{\citenamefont{Korbmacher}(2019)}]{Korbmacher}
\bibinfo{author}{\bibfnamefont{D.}~\bibnamefont{Korbmacher}}
  (\bibinfo{year}{2019}), \bibinfo{note}{manuscript in preparation}.

\bibitem[{\citenamefont{De~Boer et~al.}(1988)\citenamefont{De~Boer, Mattens,
  Boom, Miedema, and Niessen}}]{DeBoer1988}
\bibinfo{author}{\bibfnamefont{F.~R.} \bibnamefont{De~Boer}},
  \bibinfo{author}{\bibfnamefont{W.~C.~M.} \bibnamefont{Mattens}},
  \bibinfo{author}{\bibfnamefont{R.}~\bibnamefont{Boom}},
  \bibinfo{author}{\bibfnamefont{A.~R.} \bibnamefont{Miedema}},
  \bibnamefont{and} \bibinfo{author}{\bibfnamefont{A.~K.}
  \bibnamefont{Niessen}}, \emph{\bibinfo{title}{Cohesion in metals}}
  (\bibinfo{publisher}{North-Holland}, \bibinfo{year}{1988}).

\bibitem[{\citenamefont{Raji et~al.}(2009)\citenamefont{Raji, Scandolo,
  Mazzarello, Nsengiyumva, H{\"a}rting, and Britton}}]{Raji2009}
\bibinfo{author}{\bibfnamefont{A.~T.} \bibnamefont{Raji}},
  \bibinfo{author}{\bibfnamefont{S.}~\bibnamefont{Scandolo}},
  \bibinfo{author}{\bibfnamefont{R.}~\bibnamefont{Mazzarello}},
  \bibinfo{author}{\bibfnamefont{S.}~\bibnamefont{Nsengiyumva}},
  \bibinfo{author}{\bibfnamefont{M.}~\bibnamefont{H{\"a}rting}},
  \bibnamefont{and} \bibinfo{author}{\bibfnamefont{D.~T.}
  \bibnamefont{Britton}}, \bibinfo{journal}{Phil. Mag.}
  \textbf{\bibinfo{volume}{89}}, \bibinfo{pages}{1629} (\bibinfo{year}{2009}).

\bibitem[{\citenamefont{K{\"o}ppers et~al.}(1997)\citenamefont{K{\"o}ppers,
  Herzig, Friesel, and Mishin}}]{Koeppers1997}
\bibinfo{author}{\bibfnamefont{M.}~\bibnamefont{K{\"o}ppers}},
  \bibinfo{author}{\bibfnamefont{C.}~\bibnamefont{Herzig}},
  \bibinfo{author}{\bibfnamefont{M.}~\bibnamefont{Friesel}}, \bibnamefont{and}
  \bibinfo{author}{\bibfnamefont{Y.}~\bibnamefont{Mishin}},
  \bibinfo{journal}{Acta Mat.} \textbf{\bibinfo{volume}{45}},
  \bibinfo{pages}{4181} (\bibinfo{year}{1997}).

\bibitem[{\citenamefont{Tyson and Miller}(1977)}]{Tyson1977}
\bibinfo{author}{\bibfnamefont{W.~R.} \bibnamefont{Tyson}} \bibnamefont{and}
  \bibinfo{author}{\bibfnamefont{W.~A.} \bibnamefont{Miller}},
  \bibinfo{journal}{Surf. Sci.} \textbf{\bibinfo{volume}{62}},
  \bibinfo{pages}{267} (\bibinfo{year}{1977}).

\bibitem[{\citenamefont{Benoit et~al.}(2012)\citenamefont{Benoit, Tarrat, and
  Morillo}}]{Benoit2012}
\bibinfo{author}{\bibfnamefont{M.}~\bibnamefont{Benoit}},
  \bibinfo{author}{\bibfnamefont{N.}~\bibnamefont{Tarrat}}, \bibnamefont{and}
  \bibinfo{author}{\bibfnamefont{J.}~\bibnamefont{Morillo}},
  \bibinfo{journal}{Model. Simul. Mater. Sci.} \textbf{\bibinfo{volume}{21}},
  \bibinfo{pages}{015009} (\bibinfo{year}{2012}).

\bibitem[{\citenamefont{Moustafa et~al.}(2017)\citenamefont{Moustafa, Schultz,
  and Kofke}}]{Moustafa2017}
\bibinfo{author}{\bibfnamefont{S.~G.} \bibnamefont{Moustafa}},
  \bibinfo{author}{\bibfnamefont{A.~J.} \bibnamefont{Schultz}},
  \bibnamefont{and} \bibinfo{author}{\bibfnamefont{D.~A.} \bibnamefont{Kofke}},
  \bibinfo{journal}{J. Chem. Theory Comput.} \textbf{\bibinfo{volume}{13}},
  \bibinfo{pages}{825} (\bibinfo{year}{2017}).

\bibitem[{\citenamefont{Murnaghan}(1944)}]{Murnaghan1944}
\bibinfo{author}{\bibfnamefont{F.~D.} \bibnamefont{Murnaghan}},
  \bibinfo{journal}{Proc. Natl. Acad. Sci. USA} \textbf{\bibinfo{volume}{30}},
  \bibinfo{pages}{244} (\bibinfo{year}{1944}).

\bibitem[{\citenamefont{Birch}(1947)}]{Birch1947}
\bibinfo{author}{\bibfnamefont{F.}~\bibnamefont{Birch}},
  \bibinfo{journal}{Phys. Rev.} \textbf{\bibinfo{volume}{71}},
  \bibinfo{pages}{809} (\bibinfo{year}{1947}).

\bibitem[{\citenamefont{Frenkel and Ladd}(1984)}]{Frenkel1984}
\bibinfo{author}{\bibfnamefont{D.}~\bibnamefont{Frenkel}} \bibnamefont{and}
  \bibinfo{author}{\bibfnamefont{A.~J.~C.} \bibnamefont{Ladd}},
  \bibinfo{journal}{J. Chem. Phys.} \textbf{\bibinfo{volume}{81}},
  \bibinfo{pages}{3188} (\bibinfo{year}{1984}).

\bibitem[{\citenamefont{Holland}(1963)}]{Holland1963}
\bibinfo{author}{\bibfnamefont{L.~R.} \bibnamefont{Holland}},
  \bibinfo{journal}{J. Appl. Phys.} \textbf{\bibinfo{volume}{34}},
  \bibinfo{pages}{2350} (\bibinfo{year}{1963}).

\end{thebibliography}


\begin{thebibliography}{1}
\expandafter\ifx\csname natexlab\endcsname\relax\def\natexlab#1{#1}\fi
\expandafter\ifx\csname bibnamefont\endcsname\relax
  \def\bibnamefont#1{#1}\fi
\expandafter\ifx\csname bibfnamefont\endcsname\relax
  \def\bibfnamefont#1{#1}\fi
\expandafter\ifx\csname citenamefont\endcsname\relax
  \def\citenamefont#1{#1}\fi
\expandafter\ifx\csname url\endcsname\relax
  \def\url#1{\texttt{#1}}\fi
\expandafter\ifx\csname urlprefix\endcsname\relax\def\urlprefix{URL }\fi
\providecommand{\bibinfo}[2]{#2}
\providecommand{\eprint}[2][]{\url{#2}}

\bibitem[{\citenamefont{{\v{C}}{\'a}k et~al.}(2014)\citenamefont{{\v{C}}{\'a}k,
  Hammerschmidt, Rogal, Vitek, and Drautz}}]{Cak2014}
\bibinfo{author}{\bibfnamefont{M.}~\bibnamefont{{\v{C}}{\'a}k}},
  \bibinfo{author}{\bibfnamefont{T.}~\bibnamefont{Hammerschmidt}},
  \bibinfo{author}{\bibfnamefont{J.}~\bibnamefont{Rogal}},
  \bibinfo{author}{\bibfnamefont{V.}~\bibnamefont{Vitek}}, \bibnamefont{and}
  \bibinfo{author}{\bibfnamefont{R.}~\bibnamefont{Drautz}},
  \bibinfo{journal}{J. Phys.: Condens. Matter} \textbf{\bibinfo{volume}{26}},
  \bibinfo{pages}{195501} (\bibinfo{year}{2014}).

\end{thebibliography}

\end{document}

\title{Supplemental Material for: Phase transitions in titanium with an analytic bond-order potential}

\author{Alberto Ferrari}
\email{alberto.ferrari@rub.de}
\affiliation{Interdisciplinary Centre for Advanced Materials Simulation, Ruhr-Universit{\"a}t Bochum, 44801 Bochum, Germany}
\author{Malte Schr\"oder}
\affiliation{Interdisciplinary Centre for Advanced Materials Simulation, Ruhr-Universit{\"a}t Bochum, 44801 Bochum, Germany}
\author{Yury Lysogorskiy}
\affiliation{Interdisciplinary Centre for Advanced Materials Simulation, Ruhr-Universit{\"a}t Bochum, 44801 Bochum, Germany}
\author{Jutta Rogal}
\affiliation{Interdisciplinary Centre for Advanced Materials Simulation, Ruhr-Universit{\"a}t Bochum, 44801 Bochum, Germany}
\author{Matous Mrovec}
\affiliation{Interdisciplinary Centre for Advanced Materials Simulation, Ruhr-Universit{\"a}t Bochum, 44801 Bochum, Germany}
\author{Ralf Drautz}
\affiliation{Interdisciplinary Centre for Advanced Materials Simulation, Ruhr-Universit{\"a}t Bochum, 44801 Bochum, Germany}

\date{\today}

\maketitle

\section{Parameters of the BOP model}

The bond integrals have the form:
\begin{equation}
\beta\left(r\right)=a_{1}e^{-b_{1}r^{c_{1}}}+a_{2}e^{-b_{2}r^{c_{2}}}
\end{equation}
The density function in the embedding term is a third-order spline \cite{Cak2014} with two nodes:
\begin{equation}
\rho\left(r\right)=\frac{1}{2}\left[\alpha_{1}H\left(r_{1}-r\right)\left(r_{1}-r\right)^{3}+\alpha_{2}H\left(r_{2}-r\right)\left(r_{2}-r\right)^{3}\right] \quad ,
\end{equation}
where $H\left(r\right)$ is the Heavyside step function.
The repulsive two-body function is given by:
\begin{equation}
\phi\left(r\right)=Ae^{-Br^{C}} \quad .
\end{equation}
Tab.~\ref{parameters} reports the optimized values of the fitting parameters.

\begin{table*}[h!]
\begin{ruledtabular}
\begin{tabular}{ccccccc}
 & $a_{1}$ & $b_{1}$ & $c_{1}$ & $a_{2}$ & $b_{2}$ & $c_{2}$\tabularnewline
\hline 
$dd\sigma$ & -14.164826 & 1.2075490 & 0.92961725 & -3.4004649 & 0.86314510 & 1.7472806\tabularnewline
$dd\pi$ & 12.174409 & 1.2798606 & 1.0533223 & 5.4764647 & 1.1052612 & 1.3564281\tabularnewline
$dd\delta$ & -8.7311481 & 1.5381298 & 0.97829878 & 84.439899 & 3.5222294 & 0.79441562\tabularnewline
\hline 
emb.~(nodes, in \AA): & 5.1199622 & 4.6518784 & emb.~(amps., in eV$^2$): & 0.892508267 & \multicolumn{2}{c}{-1.6220065}\tabularnewline
\hline 
rep.& \multicolumn{2}{c}{$A=$12.094813} & \multicolumn{2}{c}{$B=$0.12201823} & \multicolumn{2}{c}{$C=$ 3.4464705}\tabularnewline
\end{tabular}
\end{ruledtabular}
\caption{The optimized parameters of the BOP for Ti. The amplitudes $a$ and $A$ are in eV. The inverse lengths $b$ and $B$ have units of \AA$^{-c}$ and \AA$^{-C}$ respectively. }
\label{parameters}
\end{table*}

\section{Transformation paths}

\begin{figure*}
\begin{centering} 
\includegraphics[scale=0.20]{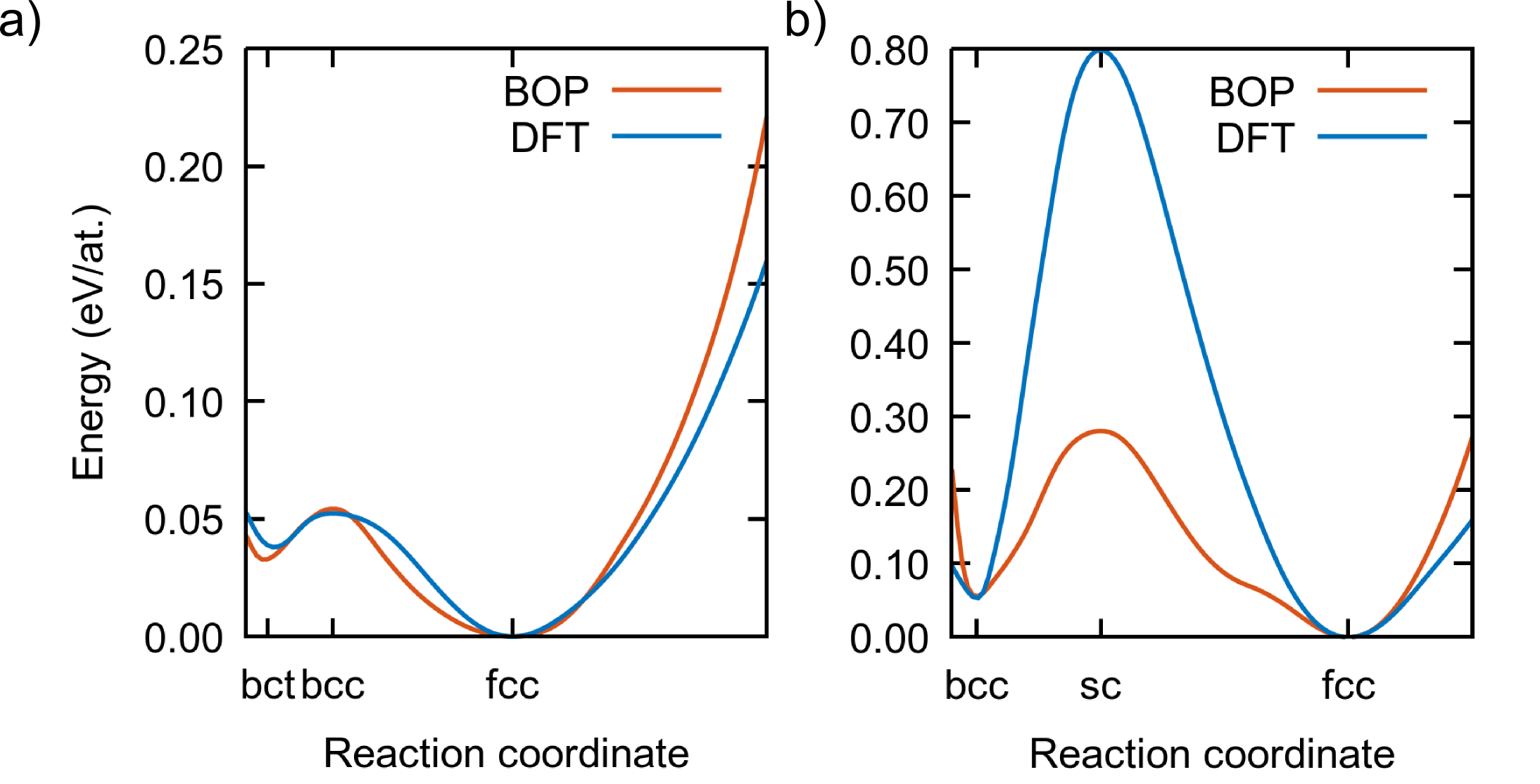}
\par\end{centering} 
\caption{\textbf{a)} Tetragonal and \textbf{b)} trigonal transformation paths calculated with BOP and DFT. The zero of energy coresponds to the lowest minimum for each path.}
\label{tr_path}
\end{figure*}

The tetragonal path shown in Fig.~\ref{tr_path} is very well described by the BOP, although the intermediate points were not considered during the parametrization.
The potential is even able to capture the shallow minimum corresponding to the body-centered tetragonal (bct) structure along the tetragonal path, completely absent from the fit set.
The energy of the simple cubic (sc) structure in the trigonal path is quite underestimated by our potential, as already discussed in the main article.

\bibliography{./bib/biblio}